\def\y{\mathbf{y}}

\def\z{\mathbf{z}}
\def\r{\mathbf{r}}
\def\x{\mathbf{x}}
\def\g{\mathbf{g}}
\def\a{{\mathbf \alpha}}
\def\e{\mathbf{e}}
\def\M{\mathbf{M}}
\def\A{\mathbf{A}}

\def\R{\mathbf{R}}

\documentclass[twoside,journal,10pt]{IEEEtran}
\usepackage{cite}

\ifCLASSINFOpdf

   \usepackage[pdftex]{graphicx}
\else
\fi
\usepackage{amsmath}
\usepackage{algorithm}
\usepackage{algorithmic}
  \usepackage[caption=false,font=footnotesize]{subfig}
\begin{document}
%
\title{Distributed Computation of Linear Inverse Problems with Application to Computed Tomography Reconstruction}
%
%
%

\author{Yushan~Gao,~
        Thomas~Blumensath,
\thanks{Manuscript received September 01, 2017. This work was supported by EPSRC grant EP/K029150/1, a University of Southampton PGR scholarship, a Faculty of Engineering and the Environment Lancaster Studentship and the China Scholarship Council.}
\thanks{Y. Gao and T. Blumensath are with the Faculty of Engineering and Environment, University of Southampton, Southampton, SO17 1BJ, UK (email: yg3n15@soton.ac.uk; Thomas.Blumensath@soton.ac.uk).
}
}

\maketitle

\begin{abstract}
The inversion of linear systems is a fundamental step in many inverse problems. Computational challenges exist when trying to invert large linear systems, where limited computing resources mean that only part of the system can be kept in computer memory at any one time. We are here motivated by tomographic inversion problems that often lead to linear inverse problems. In state of the art x-ray systems, even a standard scan can produce 4 million individual measurements and the reconstruction of x-ray attenuation profiles typically requires the estimation of a million attenuation coefficients. To deal with the large data sets encountered in real applications and to utilise modern graphics processing unit (GPU) based computing architectures, combinations of iterative reconstruction algorithms and parallel computing schemes are increasingly applied. Although both row and column action methods have been proposed to utilise parallel computing architectures, individual computations in current methods need to know either the entire set of observations or the entire set of estimated x-ray absorptions, which can be prohibitive in many realistic big data applications. We present a fully parallelizable computed tomography (CT) image reconstruction algorithm that works with arbitrary partial subsets of the data and the reconstructed volume. We further develop a non-homogeneously randomised selection criteria which guarantees that sub-matrices of the system matrix are selected more frequently if they are dense, thus maximising information flow through the algorithm. A grouped version of the algorithm is also proposed to further improve convergence speed and performance. Algorithm performance is verified experimentally.
\end{abstract}


%
\IEEEpeerreviewmaketitle

\section{Introduction}
%
%
%
%
\IEEEPARstart{I}{n} transmission computed tomography (CT), standard scan trajectories, such as rotation based or helical trajectories, allow the use of efficient analytical reconstruction techniques such as the filtered backprojection algorithm (FBP) \cite{sagara2010abdominal,hoffman1979quantitation} and the Feldkamp Davis Kress (FDK) \cite{feldkamp1984practical,rodet2004cone} method. However, in low signal to noise settings, if scan angles are under-sampled or if nonstandard trajectories are used, then less efficient, iterative reconstruction methods can provide significantly better reconstructions \cite{gervaise2012ct,wang2008outlook,deng2009parallelism,willemink2013iterative}. 
These methods model the x-ray system as a linear system of equations:
\begin{equation}
\mathbf{y} = \mathbf{Ax} + \mathbf{e},
\end{equation}
where $\mathbf{y},\mathbf{A},\mathbf{x}$ and $\mathbf{e}$ are projection data, system matrix, reconstructed image vector and measurement noise respectively. However, the relatively lower computational efficiency limits their use, especially in many x-ray tomography problems, where $\mathbf{y}$ and $\mathbf{x}$ can have millions of entries each \cite{ni2006review} and where $\mathbf{A}$, even though it is a sparse matrix, can have billions of entries. 

For realistic dataset sizes, these methods thus require significant computational resources, especially as the matrix $\mathbf{A}$ is seldom kept in memory but is instead re-computed on the fly, which can be done relatively efficiently using the latest Graphical Processor Unit (GPU) based parallel computing platforms. However, as GPUs have limited internal memory, this typically requires the data $\mathbf{y}$ and/or reconstruction volume $\mathbf{x}$ to be broken into smaller subsets on which individual computations are performed. The development of efficient algorithms that only work on subsets of the data at any one time is thus becoming of increasing interest \cite{hsieh2013recent,bilbao2004performance}.

Currently, most of these methods can be divided into two categories: \emph{row action methods}, which operate on subsets of the observations $\mathbf{y}$ at a time and \emph{column action methods}, which operate on subsets of the voxels $\mathbf{x}$ at a time \cite{censor1981row,watt1994column,elfving1980block}. 

\textbf{Row action methods}:A classical method is the Kaczmarz algorithm (also known as the Algebraic Reconstruction Technique (ART)) which together with its block based variants are widely used in tomographic image reconstruction  \cite{deng2009parallelism,kole2005parallel,li2013adaptive,censor1983strong}. Increased convergence rates are obtained with  block Kaczmarz methods, which are also known as the Simultaneous Algebraic Reconstruction Technique (SART) \cite{andersen1984simultaneous,bilbao2004performance}. By simultaneously selecting several rows of projection data, SART can also be converted into efficient parallel strategies \cite{Censor1988}. Apart from the Kaczmarz method, another classical row action method is the simultaneous iterative reconstruction technique (SIRT). The SIRT method updates each reconstructed image voxel by combining all projection values whose corresponding x-rays passing through this voxel. In this way, the convergence rate significantly increases but so does the computation load \cite{gregor2008computational,tang2012using}. To lower the computational load within one iteration, the block form of SIRT is widely used in distributed computing systems \cite{bilbao2004performance,benson2005framework,hudson1994accelerated}, yielding results of superior quality at the cost of increasing the reconstruction time. 

Since in CT systems the system matrix is often large and sparse, component averaging (CAV) and its block form (BICAV) methods have been developed to utilise the  sparsity property \cite{censor2001component,censor2002block,censor2001bicav}.  Furthermore, BICAV uses an additional coefficient matrix based on the number of non-zero elements in each column within a row block of the system matrix, thus significantly increases the convergence rate compared to the original CAV method. 

\textbf{Column action methods}: Column action methods are also called iterative coordinate descent method(ICD). They reduces the N-dimensional optimization problem into a one-dimensional problem, which is shown to have a faster convergence rate in the margins of the reconstructed image \cite{bouman1996unified}. A non-homogeneous (NH-ICD) update strategy is proposed in \cite{yu2007non,yu2011fast} to increase the convergence rate. The strategy first generates a pixel/voxel selection criterion and then based on this criterion voxels which are furthest from convergence  are selected to be updated. To increase the scalability and parallelism, a block form ICD (B-ICD) is proposed in \cite{benson2010block}. The volume object is sliced along with the helical rotation axis and within one slice several neighbouring voxels are grouped together as a block. After all blocks within one slice are updated, the slice index steps along with the axial direction to the next slice and repeats the same iteration. Experiments show that a bigger block size  is of help to increase the convergence rate, but at the cost of heavier computation amount. To reduce the computation complexity brought by B-ICD, the ABCD algorithm, which is a derivative of B-ICD method, combines the pixel line feature \cite{yu2007non,yu2011fast} and the block form \cite{benson2010block} to update the pixel line simultaneously, leading to a faster convergence rate. Inspired by NH-ICD, the NH idea is also applied to the ABCD method by more frequently updating the axial blocks which change by the largest amount during an iteration \cite{kim2012parallelizable}. However, this block method is only suitable for standard CT scanners with circular or helical scanning trajectories and may not be appropriate for arbitrary scanning trajectories.

To optimize systems of linear equation $\y=\A\x$,  \cite{elfving2016convergence} presents a concise summary of both row action and column action algorithm, which are show in Algo.\ref{Row} and Algo.\ref{Col}.
\begin{algorithm}
  \caption{Generic row action iteration}
  \label{Row}
  \begin{algorithmic}
  \STATE Initialization: $\x^0\in \R^n$ is arbitrary. Both matrix $\A$ and projection data $\y$ are divided into $p$ row blocks. $\A_i$ and $\y_i$ are corresponding $i^{th}$ row blocks. $\x^k$ is the estimate or $\x$ in the $k^{th}$ iteration. $\omega_i$ and $\M_i$ are relaxation parameters and coefficient matrices respectively. 
  \FOR {$k=0,1,2,...$(epochs or outer iterations)}
  \STATE $\z^0=\x^k$
  \FOR {$i=1,2,...,p$ (inner iterations)}
  \STATE $\z^i=\z^{i-1}+\omega_i\A_i^T\M_i(\y_i-\A_i\z^{i-1})$
  \ENDFOR
  \STATE $\x^{k+1}=\z^p$
  \ENDFOR
  \end{algorithmic}
\end{algorithm}
\begin{algorithm}
  \caption{Generic column action iteration}
   \label{Col}
  \begin{algorithmic}
  \STATE Initialization: $x^0\in \R^n$ is arbitrary. Both matrix $\A$ and vector $\x$ are divided into $q$ column blocks. $\A^j$ and $\x_j$ are corresponding $j^{th}$ column blocks.  $\r^{0,1}=\y-\A\x^0$. $\omega_j$ and $\M_j$ are relaxation parameters and coefficient matrices respectively. 
  \FOR {$k=0,1,2,...$(epochs or outer iterations)}
  \FOR {$j=1,2,...,q$(inner iterations)}
  \STATE $\x_j^{k+1}=\x_j^k+\omega_j\M_j(\A^j)^T\r^{k,j}$
  \STATE $\r^{k,j+1}=\r^{k,j}-\A^j(\x_j^{k+1}-\x_j^{k})$
  \ENDFOR
  \STATE $\r^{k+1,1}=\r^{k,q+1}$
  \ENDFOR
  \end{algorithmic}
\end{algorithm}

These two generic algorithms can be applied in parallel by parallelising the inner loop and summing or averaging the updates of $\z^i$ of $\r^{i,j}$ respectively. Each algorithm has their own advantages and drawbacks. For row action  method, each sub-iteration does not require the calculation or storage of the entire matrix $\A$ in advance but only needs to calculate a row block $\A_i$ at a time. This can be an advantage in large 3D reconstruction problems where the storage of the whole matrix is infeasible. However, if the algorithm is applied in a parallel form, then each processor (or node) needs to store the whole reconstructed image vector $\x$ as each update is performed on the entire image, which can be computationally challenging  in large data reconstructions, especially in terms of the required forward projection $\A_i\x$ and back projection $\A_i^T\r_i$ \cite{censor2001component}. 

On the other hand, using column action algorithms in a parallel computing scheme does not require each processor to store the whole reconstructed image but only to store a small part of it. However, they instead require access to all of $\y$ or $\r$, which again can be prohibitive in large scale situations. 

Thus, row action and column action methods require access to the entire vectors $\y$ (or $\r$) or $\x$ in each sub-iteration. There are few exceptions to this. One exception is the work in \cite{palenstijn2015distributed}, where a row action method is discussed in which each node  only requires parts of the reconstructed image vector $\x$. By splitting the image volume along planes perpendicular to the rotation axis, this variant of SIRT calculates the area of overlap on the detector between two adjacent sub-volumes. Only the data within the overlapping areas on the detector are exchanged between computation nodes. However, this method only works for circular scan trajectories and its scalability is limited due to the requirement that the overlap area should be small to minimise data communication overheads. 

Our goal here is to develop an algorithm working with generic x-ray tomographic scanning trajectories and with large, realistic data sizes. In this paper, we thus develop a novel algorithm called coordinate-reduced steepest gradient descent (CSGD) that combines aspects of row action and column action methods. Unlike traditional algorithms, our parallelizable algorithm operates on \emph{arbitrary} subsets of both $\x$ and $\y$ at any one time. There are thus no longer any restrictions on scan trajectories or on the way in which the  volume is decomposed. The algorithm is thus applicable to arbitrary scan trajectories and is scalable so that it can be run on a range of computing platforms, including low memory GPU clusters and high performance CPU based clusters. In our algorithm, the reconstruction volume is divided into several sub-volumes, which can be updated separately at different computing nodes. This update is based on a subset of the observations, as well as the reconstructed volume, so that only a small sub-matrix of the system matrix $\A$ is used in each step. 

The rest of this paper is organised as follows. The first section gives an overview over our new block iterative method and the second section describes the proposed algorithm in more details. Some simulation results are illustrated in section \ref{Section3} and conclusions and further discussions are presented in the last section.
\section{Description of CSGD }
\subsection{Notation}
In this paper we assume a linear, i.e. monochromatic x-ray model \cite{soleimani2015introduction,guo2016convergence,beister2012iterative}: 
\begin{equation}
\y=\A\x +\e.
\end{equation} 
For now, we use a simple least square cost function to solve this inverse problem:
\begin{equation}
f(\x)=(\y-\A\x)^T(\y-\A\x),
\end{equation}
where $\A$ is the system matrix derived using, for example, Siddon's method \cite{jacobs1998fast}. The element $A_i^j$ is the intersection length of the $i^{th}$ x-ray beam with the $j^{th}$ reconstruction voxel. In this paper,  $I$ will be an index set that indexes rows in $\A$ (or the subset of x-ray measurements $\y$)  $I={i_1,i_2,...}$. Similarly, $J$ will be a set of column indexes of $\A$ (or the index set of a subset of voxels $\x$)  $J={j_1,j_2,...}$. Thus the matrix $\A_{I_i}^{J_j}$ will be a sub-matrix of $\A$ with row indexes $I_i$ and column indexes $J_j$. Thus we can divide the linear system into several blocks:
\begin{equation}
\begin{bmatrix}
\y_{I1}\\
...\\
\y_{Im} 
\end{bmatrix}\approx
\begin{bmatrix}
\A_{I_1}^{J_1} \quad \A_{I_1}^{J_2} \quad ... \quad \A_{I_1}^{J_n}\\
... \\
\A_{I_m}^{J_1} \quad \A_{I_m}^{J_2} \quad ...
\quad \A_{I_m}^{J_n}
\end{bmatrix}
\begin{bmatrix}
\x_{J1}\\
... \\
\x_{Jn}
\end{bmatrix}\equiv
\begin{bmatrix}
\A_{I_1}\\
... \\
\A_{I_m}
\end{bmatrix}\x.
\label{e1}
\end{equation}
Note that the index sets can be arbitrary partitions of the columns and rows and do not necessarily have to be consecutive. For the convenience of the latter discussion, we also define $\r=\y-\A\x$ and $\r_I$ as the subset of $\r$ defined as $\y_I-\A_I\x$.
\subsection{Derivation of the algorithm}
The key idea in CSGD is to minimize the partial residual $\r_I$ with partial coordinate $\x_J$. In one iteration, after selecting row and column index sets $I$ and $J$, the object function becomes:
\begin{equation}
f(\x)=\r_I^T\r_I=(\y_I-\A_I\x)^T(\y_I-\A_I\x).
\end{equation}
The steepest descend direction $\g$ of $f(\x)$ is then:
\begin{equation}
\g=-\nabla f(x)=\A_I^T(\y_I-\A_I\x)=\A_I^T\r_I.
\end{equation}
Assume that only those voxels whose indices are in the set $J$ are updated, then the new descend direction becomes:
\begin{equation}
\label{direction}
\g=\begin{bmatrix}
\g_J\\
\mathbf{0}
\end{bmatrix}
=\begin{bmatrix}
(\A_I^J)^T\r_I\\
\mathbf{0}
\end{bmatrix}
\end{equation}
and the update on the selected voxels becomes:
\begin{eqnarray}
\x_J^{n+1}&=&\x_J^{n}+\mu \g_J,\\ 
\x_{\hat{J}}^{n+1}&=&\x_{\hat{J}}^{n}.
\end{eqnarray}
where $\mu$ is the gradient step length and $\hat{J}$ is the complement to the set $J$.

Ideally, we would like to compute the step length $\mu$ such that $f(\x^{n+1})$ is minimised:
\begin{equation}
\nabla f(\x^{n+1})^T\g=0,
\end{equation} 
where $\nabla f(\x^{n+1})=(\A_I)^T(\A_I\x^{n+1}-\y_I)$. Using the fact that $\A_I\g=\A_I^J\g_J$, the steepest step length is thus calculated as:
\begin{equation}
\label{step}
\mu=\frac{{\g_J}^T(\A_I^J)^T\r_I}{{\g_J}^T(\A_I^J)^T\A_I^J\g_J}=\frac{{\g_J}^T\g_J}{{\g_J}^T(\A_I^J)^T\A_I^J\g_J}.
\end{equation}
With this step length, each iteration only requires the information of $\A_I^J$, which significantly reduces the calculation amount when the matrix is generated on the fly with Siddon's method. Although the calculation of the step size $\mu$ still requires access to the error $\r_I=\y_I-\A_I\x$, the calculation of $\r_I$ can be replaced by an update process \cite{shen2016accelerated,qu2016coordinate}. Since the update within one iteration only changes $\x_J$, we have:
\begin{equation}
\A_I\x^{n+1}=\A_I\x^n-\A_I^J\x_J^n+\A_I^J\x_J^{n+1},
\end{equation}
where $\x_J^n$ is the result of the $n^{th}$ iteration.
So the update on $\r_I$ can be written as:
\begin{equation}
\r_I^{n+1}=\r_I^n+\A_I^J(\x_J^n-\x_J^{n+1}) = \r_I^n-\mu \A_I^J(\g_J).
\end{equation}
We can see that this update only requires computation and storage of $\A_I^J\x_J^{n+1}$. 

An important issue is that the step size derived in (\ref{step}) minimizes $\|\r_I\|$ instead of $\|\r\|$ and that the step length is always positive. However, we are not interested in the reduction of $\r_I$ but in the reduction of $\r$. Our choice of $\mu$ can thus potentially be too large to reduce $\r$. Furthermore, it is not guaranteed that our update direction $\g$ is always a descend direction. To stabilise our algorithm, we thus introduce an additional relaxation parameter $\beta$ into the calculation of $\mu$. This helps us to avoid overshooting the minimum if $\g$ is a descend direction whilst in cases in which $\g$ is not a descend direction, the increase in $r$ remains small. The pseudo-code of the basic computation blocks is shown in Algo.\ref{alg1}.

 \begin{algorithm}
  \caption{The algorithm for a basic iteration}
  \label{alg1}
  \begin{algorithmic}
  \STATE Initialization: select system matrix's row index $I$ and column index $J$
  \STATE $\g_J=(\A_{I}^J)^T\r_I$
  \STATE $\mu = \beta \frac{(\g_J)^T\g_J}{(\g_J)^T(\A_{I}^J)^T\A_{I}^J\g_J}$
  \STATE $\x_J=\x_J+\mu \g_J$
  \STATE $\z_I^{J}=\A_I^J\x_J$
  \end{algorithmic}
\end{algorithm}
It is straightforward to see that these computations can be computed in parallel over $J$ since parameters in the update on $\x_J$ are independent of each other. Further analyses and improvements show that the parallel computation can be even performed over both subsets of $J$ and $I$.

The main trick in parallelizing the above code is to estimate the error $\r$. Ideally, after the parallel computations have updated subsections of the volume $\x$, we would need to compute a new error vector $\r$, which is required in the computation of subsequent gradient directions. However, this would require the computation of $\r_I^n+\A_I^J(\x_J^n-\x_J^{n+1})$. We instead use a scheme that approximates $\r$. This is done by calculating  vectors $\z_I^{J,n+1}=\A_I^J\x_J^{n+1}$. 

Algo.\ref{alg1b} shows a fully parallel computing scheme over both row and column blocks. Different sub-matrix block of matrix $A$ and sub-volume slice $x_J$ are assigned to calculating nodes simultaneously and the master node communicate with all computing nodes by averaging over the individual updates of subsets $\x_J$. It is worth noting at this point that these algorithms do not require us to use all subsets $I$ and $J$ in each iteration, but also work if we randomly choose new subsets of these sets in each iteration.
 \begin{algorithm}  
  \caption{CSGD method}
   \label{alg1b}  
    \begin{algorithmic}  
  \STATE Initialization: Partition row and column indices into sets $\{I_i\}$ ($i\in [1, m]$) and $\{J_j\}$ ($j\in [1,n]$), $\x^0=\textbf{0}$, $\z_{I_i}^{j}=\mathbf{0}$ for all blocks $I_i,J_j$ and $\r=\y$.
  \FOR{epoch=1,2,...}
   \STATE $\hat{\x}=\mathbf{0}$  
     \FOR{$k=1,2,..n$}
     \STATE select $J_k$ as index $J$
   \FOR{ $i$=1,2,..$m$}
   \STATE select $I_i$ as index $I$
     \STATE $\g_J=(\A_{I}^J)^T\r_I$
     \STATE $\mu = \beta \frac{(\g_J)^T\g_J}{(\g_J)^T(\A_{I}^J)^T\A_{I}^J\g_J}$
     \STATE $\hat{\x}_J=\hat{\x}_J+\x_J+\mu \g_J$
     \STATE $\z_I^j=\A_I^J(\x_J+\mu \g_J)$ 
     \ENDFOR
     \STATE $\r = \y-\sum_j \z^j$
   \ENDFOR
   \STATE for all blocks $J$ that have been updated, $\x_J=\hat{\x}_J$/(number of times block $J$ has been updated).
  \ENDFOR 
\end{algorithmic}  
\end{algorithm}
\subsection{The importance sampling strategy}
Based on the proposed algorithm, it is straightforward to develop a random sampling strategy that goes through all projection views and all detector sub-areas arbitrarily. 
Looking at Algo.\ref{alg1b}, we could randomly select subsets of $\{I_i\}$ and $\{J_j\}$, with each set being chosen with equal probability. Considering the sparsity of $\A$ and taking inspiration from the randomized Kaczmarz method in \cite{strohmer2009randomized}, we instead develop an importance sampling strategy. Sets in $\{I_i\}$ and $\{J_j\}$ are selected with a probability that is proportional to the sparsity of the sub-matrix $\A_I^J$. To estimate this sparsity without the need to construct the full matrix $\A$,  we instead compute the overlap between the detector area and the projection of the voxels labelled by $J$. Using the importance sampling strategy for $I$ and iterating over partial $J$, Algo.\ref{alg1b} become Algo.\ref{alg3}.
 \begin{algorithm}  
  \caption{CSGD method with importance sampling}
   \label{alg3}  
    \begin{algorithmic}  
  \STATE Initialization: Partition row and column indices into sets $\{I_i\}$ ($i\in [1, m]$) and $\{J_j\}$ ($j\in [1,n]$), $\x^0=\textbf{0}$. $\z_I^j=\mathbf{0}$ for all row blocks $I_i$ and $J_j$, $\r=\y$. $\gamma$ is the percentage of selected volume blocks in the total volume blocks. $\alpha$ is the percentage of selected row blocks in the total row blocks. Relaxation parameter $\beta$ is defined as $b*P_S/P_T$, where $P_S$ is the projection area of the selected sub-volume on the selected sub-area on current iteration and $P_T$ is the total projection area for the selected volume object on the whole detector plane under the whole scanning trajectory.
  \STATE For each pair of indices from $\{I_i\}$ and {$\{J_j\}$} compute the probability $P(I,J)=P_S/P_T$ between the projection of the volume $\x_J$ and the subset of all detector pixels indexed by $I$. 
  \FOR{epoch=1,2,...}
   \STATE $\hat{\x}=\mathbf{0}$  
   \FOR{ $k=1,2,..\gamma n$}
   \STATE select a column block $J$ from {$\{J_j\}$} randomly
   \FOR{$l=1,2,...\a m$}
   \STATE select a row block $I$ from {$\{I_i\}$} without replacement with probability $P(I,J)$
     \STATE $\g_J=(\A_{I}^J)^T\r_I$
     \STATE $\mu = \beta \frac{(\g_J)^T\g_J}{(\g_J)^T(\A_{I}^J)^T\A_{I}^J\g_J}$
     \STATE $\hat{\x}_J=\hat{\x}_J+\x_J+\mu \g_J$
     \STATE $\z_I^j=\A_I^J(\x_J+\mu \g_J)$ 
     \ENDFOR
     \STATE $\r = \y-\sum_j \z^j$
   \ENDFOR
   \STATE for all blocks $J$ that have been updated, $\x_J=\hat{\x}_J$/(number of times block $J$ has been updated)
  \ENDFOR 
\end{algorithmic}  
\end{algorithm}
It should be noted that although in Algo.\ref{alg1b} and \ref{alg3} the update on $\r$ is performed as a whole, in effect, only elements which have been chosen for the latest iteration have been changed. As a result, we can also take the union of all selected $I$s and call this set $It$. The update of $\r$ then becomes $\r_{It}=\y_{It}-\sum_j \z_{It}^j$.
\subsection{Domain decompositions in tomography.}
\begin{figure}[!htb]
\centering     
\subfloat[]{\label{f2a}\includegraphics[width=1.7in]{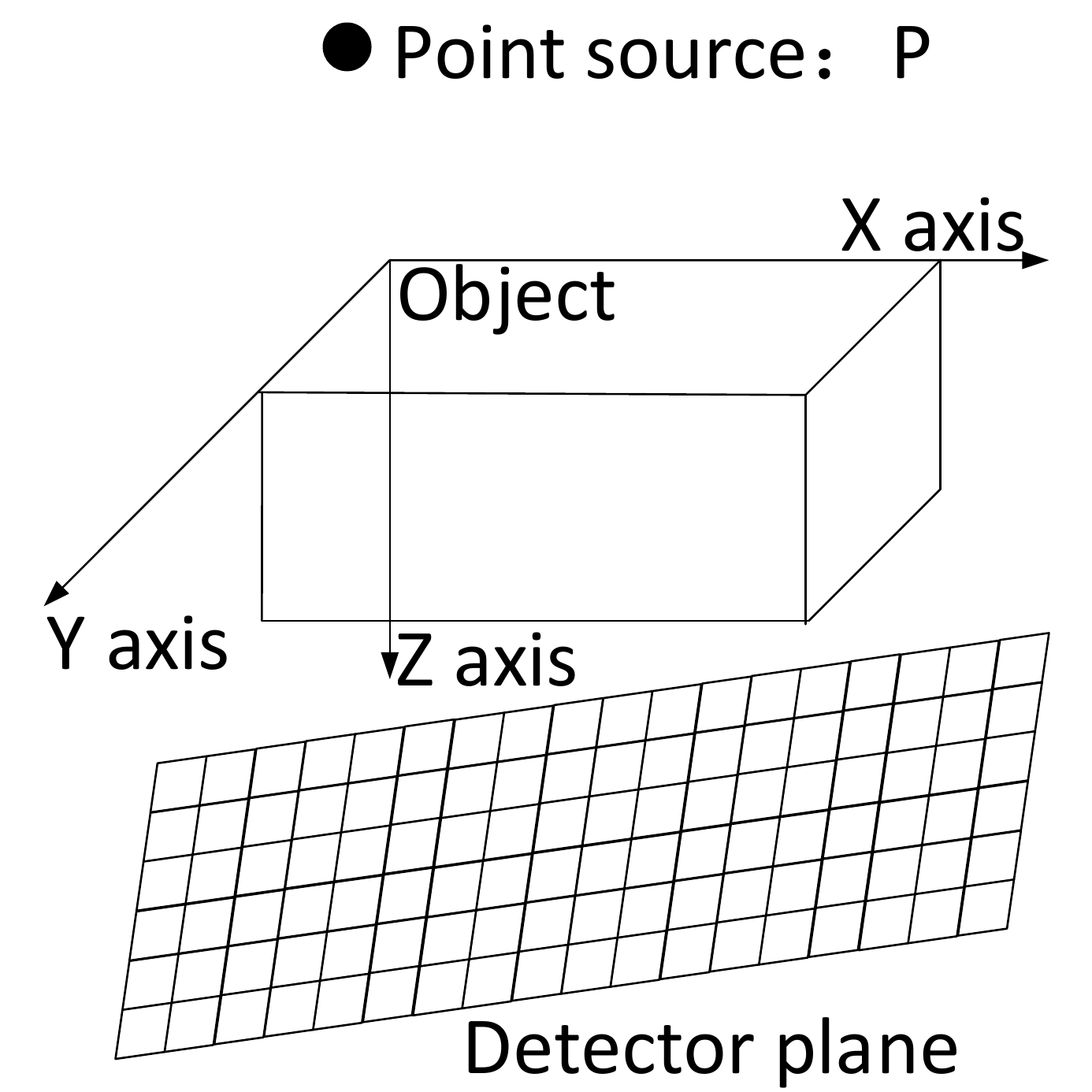}}
\subfloat[]{\label{f2b}\includegraphics[width=1.7in]{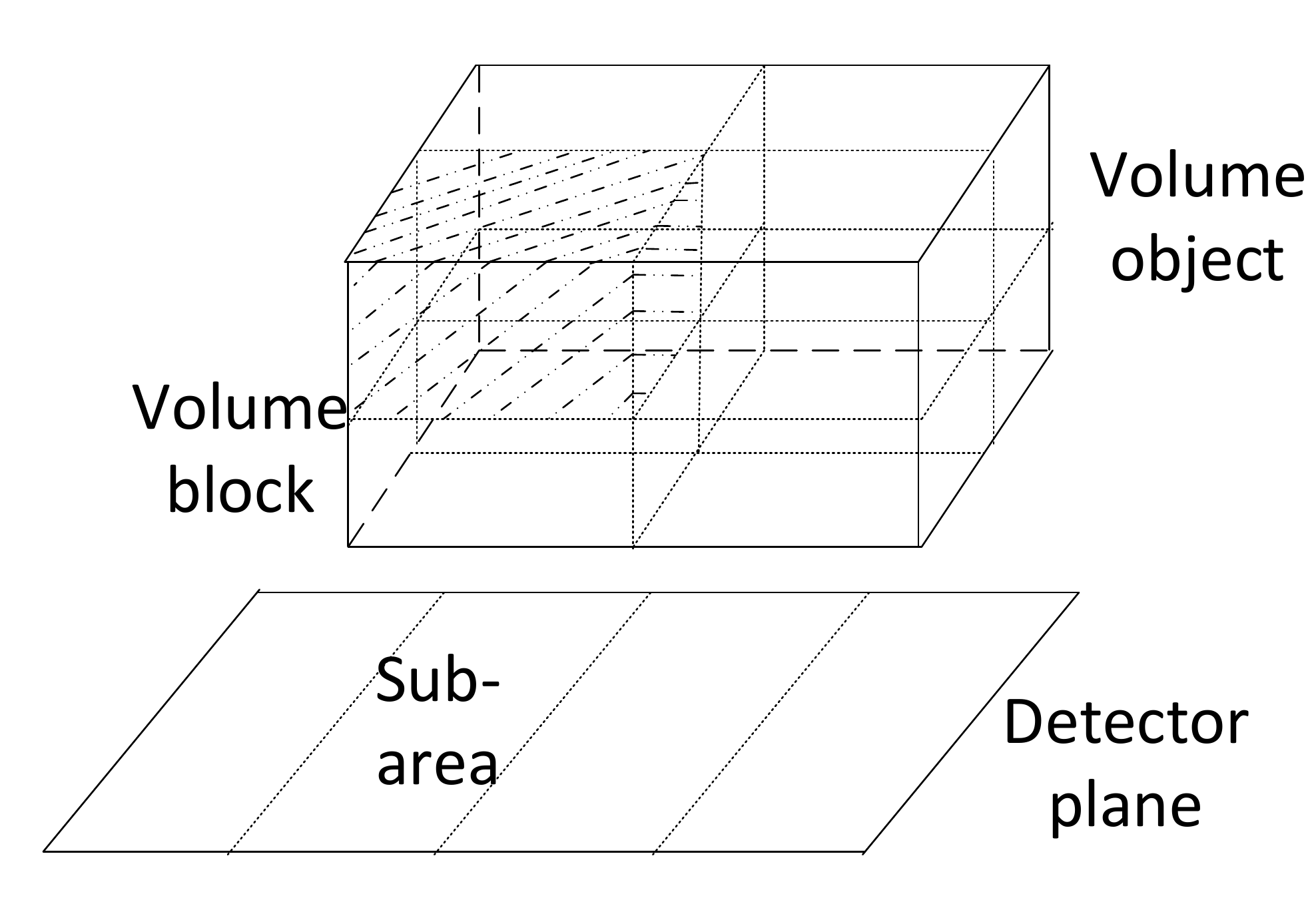}}
\caption{(a) is the geometry of a 3D scanning model and (b) is a partition method on both reconstruction volume and detector area. }
\label{f2}
\end{figure}

The above algorithms have been designed so that each computation is carried out on a generic subset $I$ (i.e. a subset of the observations) and a generic subset $J$ (i.e. a subset of the voxels) at any one time. For tomographic reconstruction, the question thus arises how to partition the observations and the reconstruction volume. Whilst generic partitions are possible, given the need to compute $P(I,J)$, it makes sense to partition the reconstruction volume and detector areas into blocks. We use a 3D cone beam CT geometry to demonstrate and similar arguments can be made for a parallel beam setup. The reconstruction volume and one pair of source/detector locations are shown in Fig.\ref{f2a}. For simplicity, the detector plane is always perpendicular to the line connecting point source and the geometry center of the detector plane.

\begin{figure}
\centering     
\subfloat[]{\label{f3a}\includegraphics[width=1.7in]{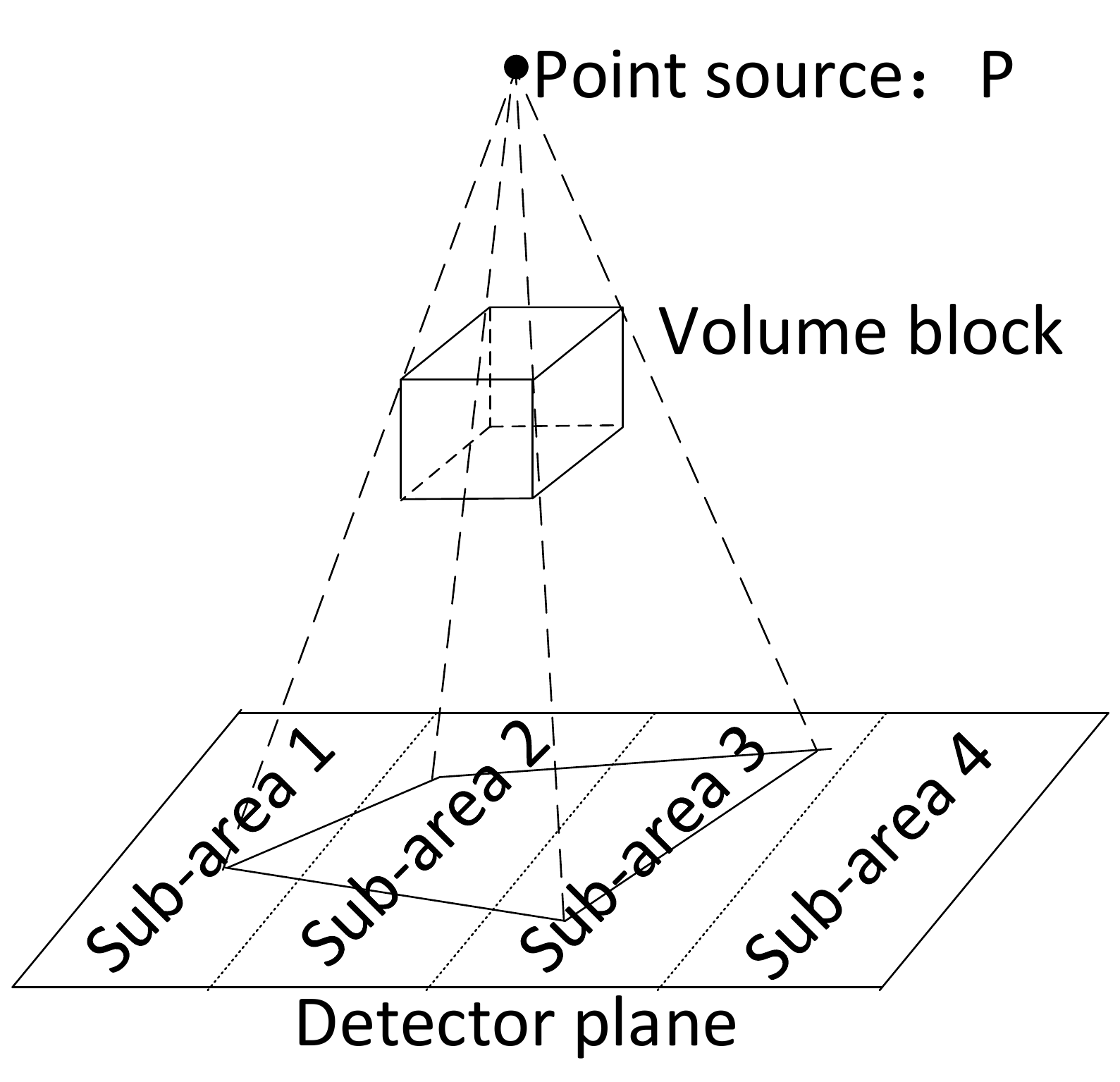}}
\subfloat[]{\label{f3b}\includegraphics[width=1.7in]{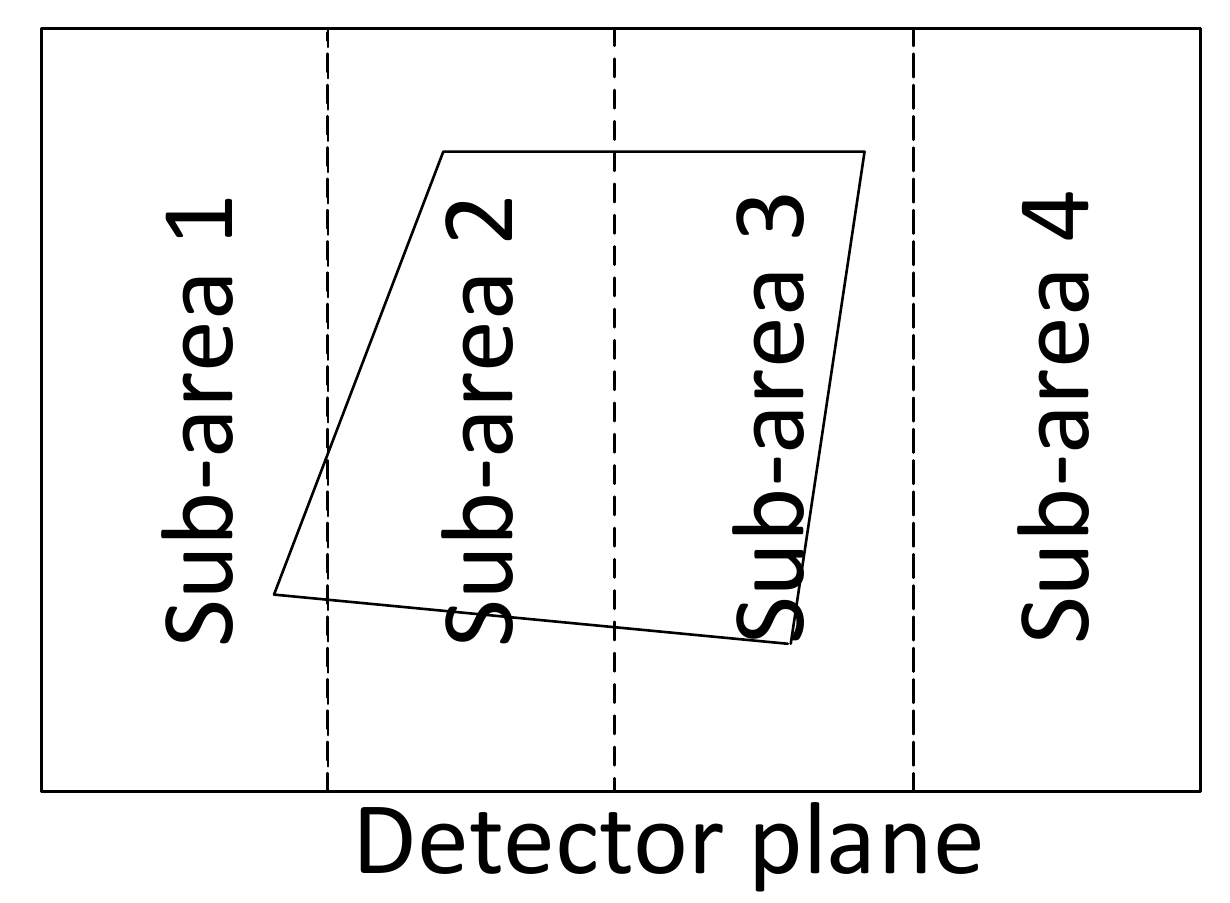}}
\caption{(a) shows a 3D model of the cone beam setup with one block of the volume being projected on a detector plane. (b) shows the projection area of the volume block, whose projection is unevenly distributed on each of sub-area.}
\label{f3}
\end{figure}

We label each location of the point source $S$ and detector location with parameter $\theta$, whose trajectories do not have to be circular or helical. For each source/detector location $\theta$, we partition the detector into blocks. We also partition the reconstruction volume into rectangular cuboids, as shown in Fig.\ref{f2b}.
For such a partition, given any one block of the reconstruction volume $J$, for each source location $\theta$, there might only be a part of the detector area that is involved in the update in CSGD. Thus, for a given $J$, the sub-matrices $\A_I^J$ can have different levels of sparsity for different $I$. An illustration of this is shown in Fig.\ref{f3}, which shows that in this fixed view, sub-area 4 on the detector does not receive rays passing through the selected sub-volume and so there is no need to select this area to update $\x_J$. What is more, the bulk of the volume block projection mainly lies in sub-areas 2 and 3 of the detector so that the corresponding $\A_{Isub2}^J$ and $\A_{Isub3}^J$ are much denser than $\A_{Isub1}^J$. This suggests that the algorithm should select sub-area 2 and 3 more frequently than sub-area 1 and this is here achieved with our importance sampling strategy.

\subsection{Group CSGD}
The advantage of having dense sub-matrices leads to the following group version of our algorithms. The idea here is to dynamically build larger, dense sub-matrices $\A_I^J$ out of a large selection of smaller sub-matrices.   In the previous CSGD method, one sub-block of the image combines only one sub-detector area for one projection view. In GCSGD, 
the sub-block combines a group of sub-areas for several projection views. It is straightforward that the GCSGD method uses more row information than the CSGD method.  Let us demonstrate the idea by modifying Algo.\ref{alg3} into Algo.\ref{alg4}.
  \begin{algorithm}  
  \caption{GCSGD method with importance sampling}
   \label{alg4}  
    \begin{algorithmic}  
  \STATE Initialization: Group size is set as $s$ while the other initialization term is the same as Algo.\ref{alg3}.
  \FOR{epoch=1,2,...}
   \STATE $\hat{\x}=\mathbf{0}$  
   \FOR{ $k=1,2,..\gamma n$}
   \STATE select a column block $J$ from {$\{J_j\}$} randomly
\STATE $l=0$
   \WHILE{$l<\a m$}
   \STATE $I_g=\mathbf{0}$
   \STATE $pin=0$
  \WHILE {$pin<s$}
   \STATE $pin=pin+1$
    \STATE $l=l+1$
   \STATE select a row block $I$ from {$\{I_i\}$} without replacement with probability proportional to $P(I,J)$
   \STATE $I_g=I_g\cup I$
   \ENDWHILE
     \STATE $\g_J=(\A_{I_g}^J)^T\r_{I_g}$
     \STATE $\mu = \beta \frac{(\g_J)^T\g_J}{(\g_J)^T(\A_{I_g}^J)^T\A_{I_g}^J\g_J}$
     \STATE $\hat{\x}_J=\hat{\x}_J+\x_J+\mu \g_J$
     \STATE $\z_{I_g}^j=\A_{I_g}^J(\x_J+\mu \g_J)$ 
     \ENDWHILE
     \STATE $\r = \y-\sum_j \z^j$
   \ENDFOR
   \STATE for all blocks $J$ that have been updated, $\x_J=\hat{\x}_J$/(number of times block $J$ has been updated).
  \ENDFOR 
\end{algorithmic}  
\end{algorithm}
\subsection{Computational complexity}
There are several important aspects when comparing computational efficiency of the methods. The methods are designed to allow parallel computation. We envisage this to be performed in a distributed network of computing nodes\footnote{A serial version running on a single computing node where each computation is done independently, but one after the other,  is also possible and this is how many of the simulations reported here were computed.}. Most of these nodes will be used to perform the parallel computations. They produce two outputs, 
\begin{enumerate}
\item $\hat{\x}_J(i) = \x_J^n+\mu\g_J$ 
\item $\z_I^j=\A_I^J\x_J$. %
\end{enumerate}
These are then either sent to larger, but slow storage or directly to other nodes, where they are eventually used to compute 
\begin{enumerate}
\item $\x_J$ = mean$_i(\hat{\x}_J(i))$ 
\item $\r = \y-\sum_j \z^j$ or $\r_{It} = \y_{It}-\sum_j \z_{It}^j$,
\end{enumerate}
which can be performed efficiently using message passing interface reduction methods.

The three main points that affect performance of the method are: 
\begin{enumerate}
\item Computational complexity in terms of multiply add operations.
\item Data transfer requirements between data storage and a processing unit as well as between different processing units.
\item Data storage requirements, both in terms of fast access RAM and in slower access (e.g. disk based) data storage.
\end{enumerate}
Each of these costs are dominated by different aspects:
\begin{enumerate}
\item Computational complexity is dominated by the computation of matrix vector products involving $\A_I^J$ and its transpose, especially as $\A$ is not generally stored but might have to be re-computed every time it is needed.  The computational complexity is thus $O(|I|*|J|)$, though computations performed on highly parallel architectures, such as modern GPUs, are able to perform millions of these computations in parallel. 
\item Data transfer requirements are dominated by the need for each of the parallel computing nodes to need $\r_I$ and $\x_J$ as input and $\hat{\x}_J(i)$ and $\z_I^j$ as output. Note that the size of the required input and output vectors are the same, the data transfer requirement is thus $O(|I|+|J|)$.
\item Central data storage requirements are dominated by the need to store the original data and the current estimate of $\x$. We also need to compute and store averages over $\hat{\x}_J(i)$ and $\z_I^j$. These computations can be performed efficiently using parallel data reduction techniques.  Our approach would mean that each node would thus require $O(|I|+|J|)$ local memory.
\end{enumerate}
\section{Simulations}
\label{Section3}
Before introducing the simulation results, We first introduce three important parameters:

$\a$: is the percentage of selected row blocks (detector areas) for one column block (volume block). 

$\beta$: is the relaxation parameter tuning the update on $x_J$. It is  expressed as $\beta=b*P_S/P_T$ as shown in Algo.\ref{alg3}. In the following simulations we mainly  change the value of $b$ to tune the relaxation parameter $\beta$.

$\gamma$: is the percentage of the selected column blocks (volume blocks) during one iteration. In our simulations,  $\gamma$ is usually set as 1, i.e. all column blocks are selected during one iteration. 

In simulations, we used two evaluation criteria to evaluate the reconstruction quality: Signal to Noise ratio (SNR):$20*log_{10}\| \x_{true}\|/\| \x_{true}-\x_{est}\|$ and observation gap:$20*log_{10}\|\y\|/\|\y-\A\x_{est}\|$, where $\y$ are the projection data, $\x_{true}$ and $\x_{est}$ are the true phantom image vector and the reconstructed image vector respectively.

\begin{figure}
\centering     
\subfloat[]{\label{f1a}\includegraphics[width=1.7in]{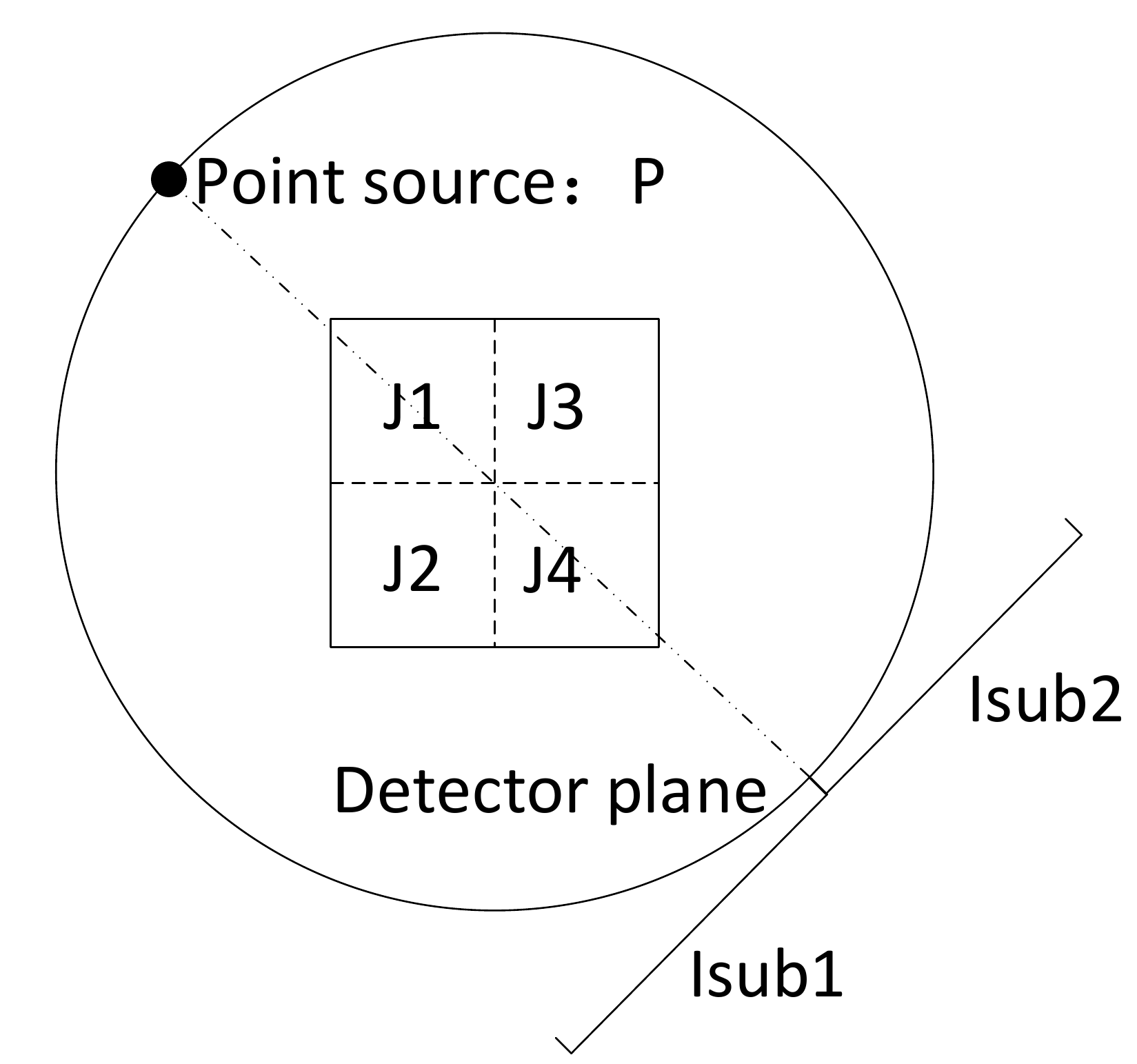}}
\subfloat[]{\label{f1b}\includegraphics[width=1.7in]{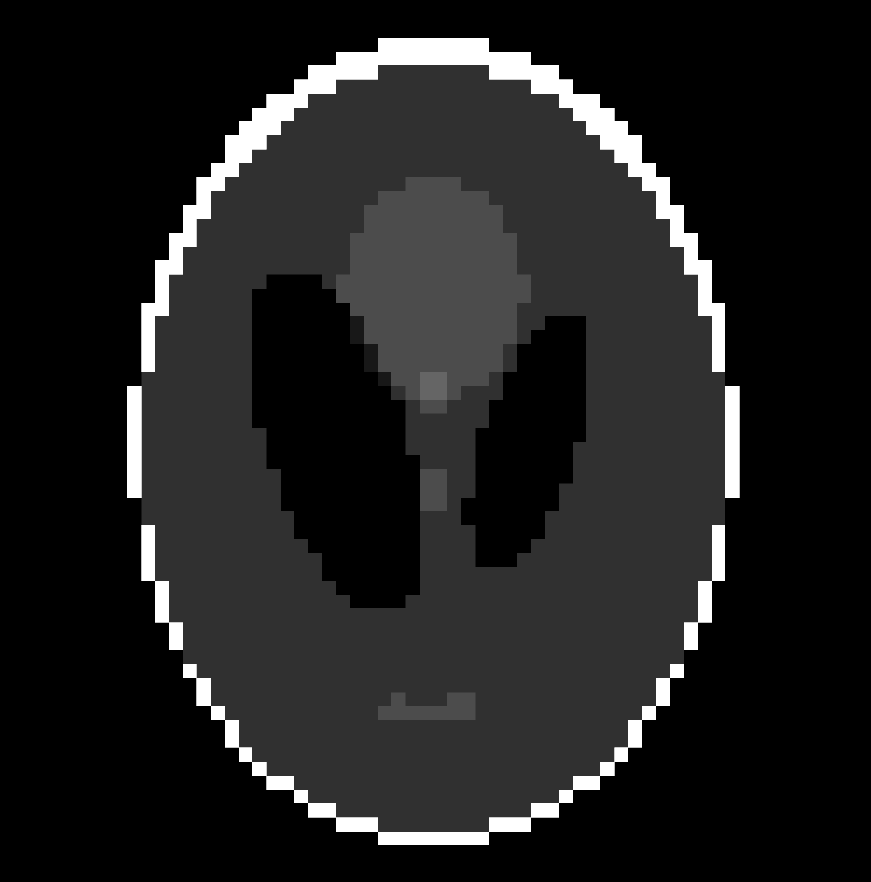}}
\caption{Basic simulation settings. (a) shows the scanning geometry and partition model in the 2D simulation. The object is partitioned from $J_1$ to $J_4$ and the detector is partitioned to two parts $Isub_1$ and $Isub_2$. The total 187 detector pixels are assigned to these two sub areas without overlapping. (b) shows the original image to be scanned and reconstructed.}
\label{f14}
\end{figure}

We explored the performance of Algo.\ref{alg4} on a range of tomographic reconstruction problems. We started with a simulated 2D phantom with 64*64 pixels. Pixel sizes were normailsed to be 1. The point source adopted a circular trajectory with radius of 115 and the rotation centre was located at the centre of the object. The linear detector had 187 pixels whose spacing was 1 and the detector was always perpendicular to the line connecting the point source and the geometric centre of the linear detector. The detector centre also followed a circular trajectory with the same radius as the source. The number of projections was 360 with the angular intervals being 1$^\circ$. The object was partitioned into 4 parts (2 parts in both vertical and horizontal directions) and the detector area was partitioned into 2 parts by default. Scan geometry including the partition method as well as the original phantom are shown in Fig.\ref{f14}. We used the parallel computing toolbox (version 6.8) in Matlab R2016a to perform CSGD as described in Algo.\ref{alg3} and GCSGD in Algo.\ref{alg4}. In our simulation, the term \emph{epoch} refers to the outer iteration. The number of sub-matrices $\A_I^J$ that the algorithm used per epoch was proportional to the parameter $\a$, where $\a=1$ means that we used all sub-matrices of $\A$, whilst $\a=0.5$ means that only half of the sub-matrices were used. As the computational load is dominated by matrix vector products, when we compared the difference in convergence rate for methods using different $\a$s (for example, $\a=1$ and $\a=0.5$), we took account of the reduction in computational effort when using smaller $\a$ by scaling the epoch count by multiplying it by $\a$, a measure we call \emph{effective epoch}. 

As mentioned before, CSGD and GCSGD methods can be applied in both using importance sampling and random sampling strategies. In our simulations, unless it is particularly mentioned, the sampling strategy always adopts the importance one.

We run CSGD and GCSGD algorithms for over one thousand iterations and show their average convergence and stability. However, in realistic applications we are typically only interested in the initial iterations. For example, the convergence performance within 50 effective epochs. This is because in real applications, due to the large data sets that have to be processed and due to the influence of noise and model missmatch, performing more iterations is typically not feasible and is often not helpful in obtaining better reconstruction results.  
\subsection{Influence of $\beta$}
\begin{figure}[!]
\centering
\includegraphics[width=3.5in]{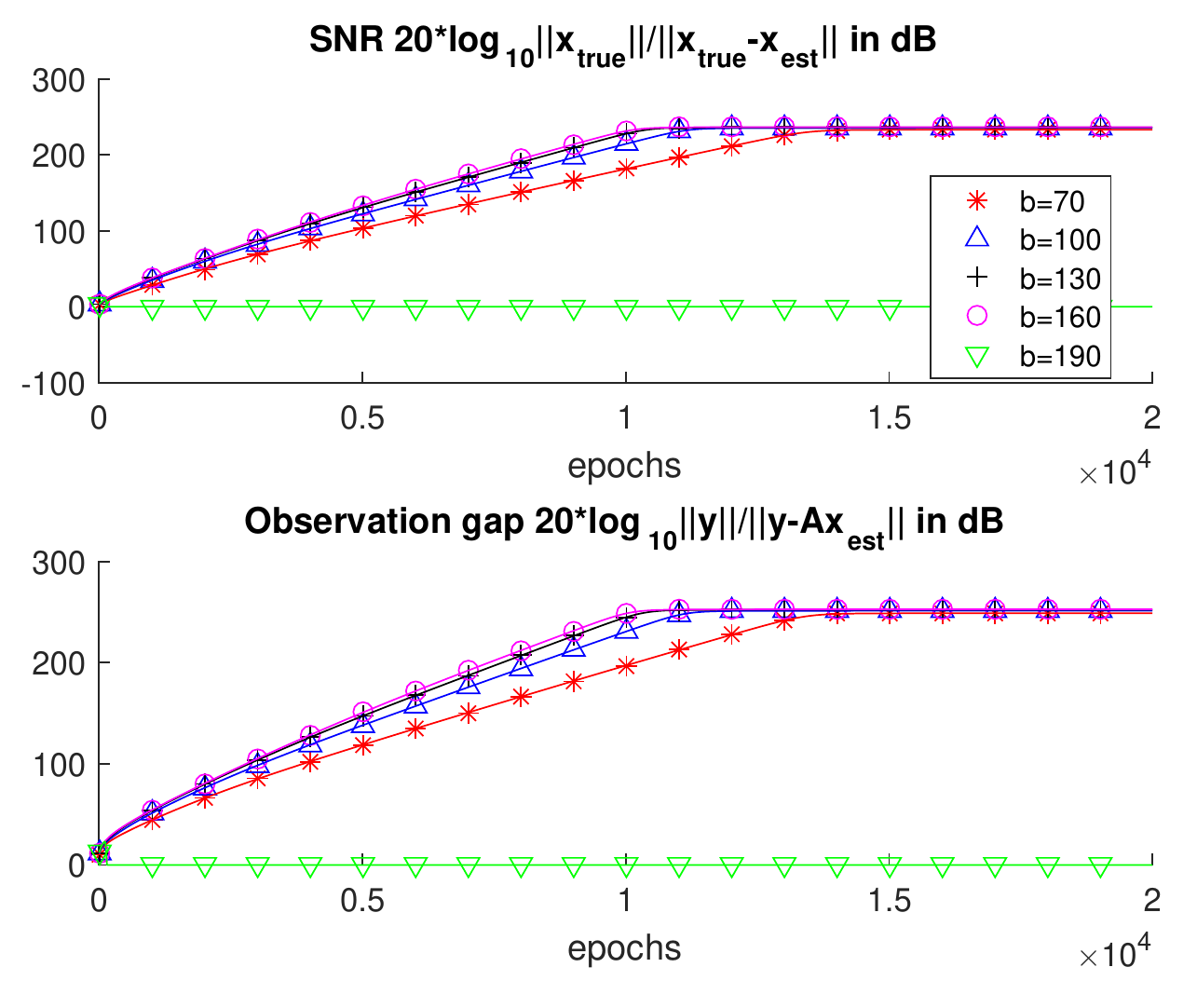}
\caption{Different $\beta$ lead to different convergence rates when reconstructing the 2D image. $\a$ and group size are both set as 1. All 4 volume sub-blocks are selected in each epoch.}
\label{f4}
\end{figure}

We started with an evaluation of the optimal choice of $\beta$, which is crucial for the performance of the method. A simulation was conducted to show the difference in convergence rate when changing $b$. Results are shown in Fig.\ref{f4}. 

It can be seen that from $b=70$ to $b=160$, larger $\beta$ increases the convergence rate. However, further increasing the value of $b$ to 190 leads to a divergence of the algorithm, which suggests that there is a range for $\beta$ that is guranteed to converge. However, currently the justification for our choice of $\beta$ remains empirical and more analysis of the algorithm is required to fully understand the optimal parameter.  

\subsection{Choice of $\alpha$.}
\begin{figure}[H]
\centering
\includegraphics[width=3.5in]{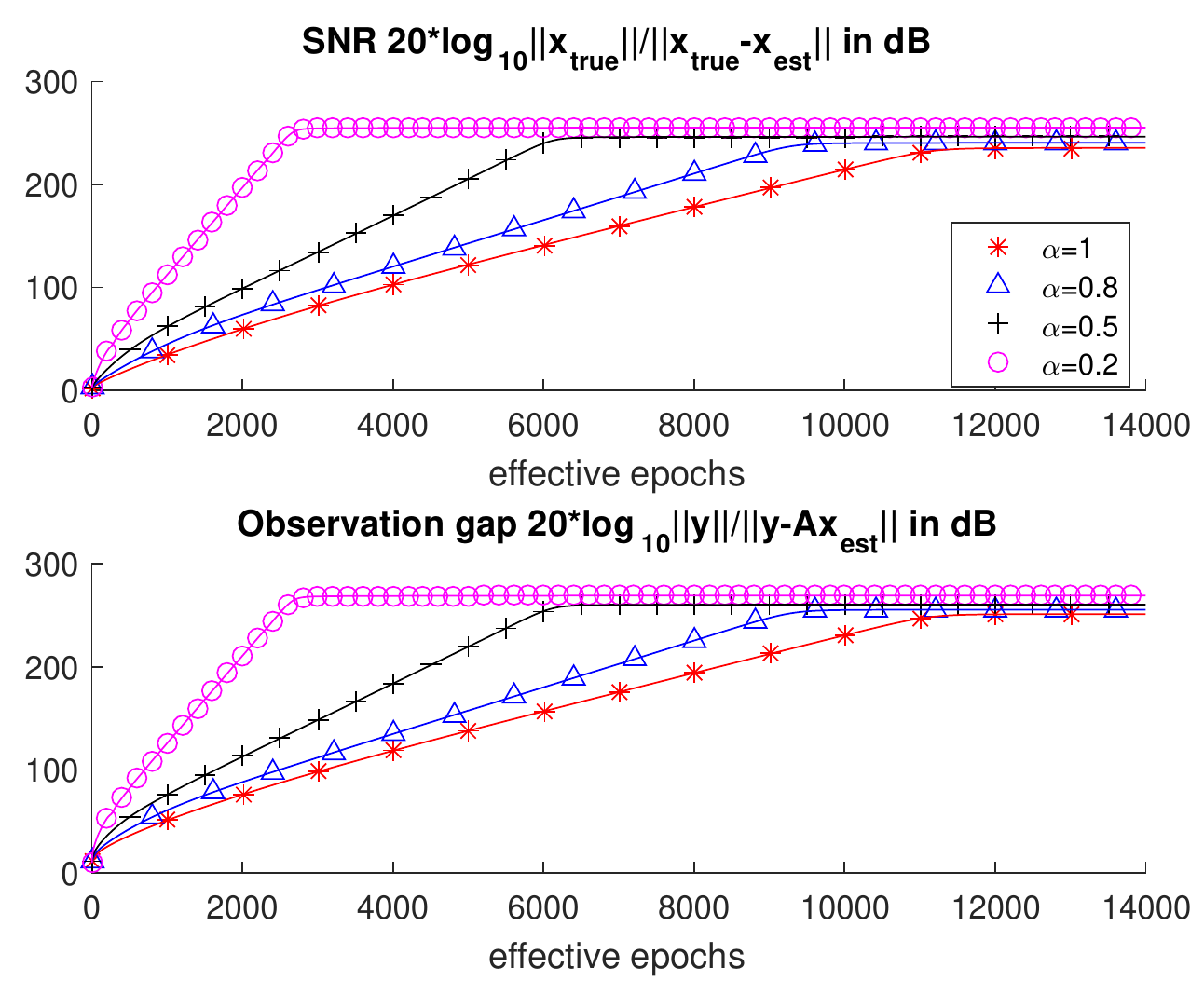}
\caption{When $b=100$, groups size is 1 and all 4 volume blocks are selected for each epoch, setting $\a$ smaller than 1 is guaranteed to achieve high precision level and is of help to increase the convergence rate.}
\label{f15}
\end{figure}

\begin{figure}[H]
\centering
\includegraphics[width=3.5in]{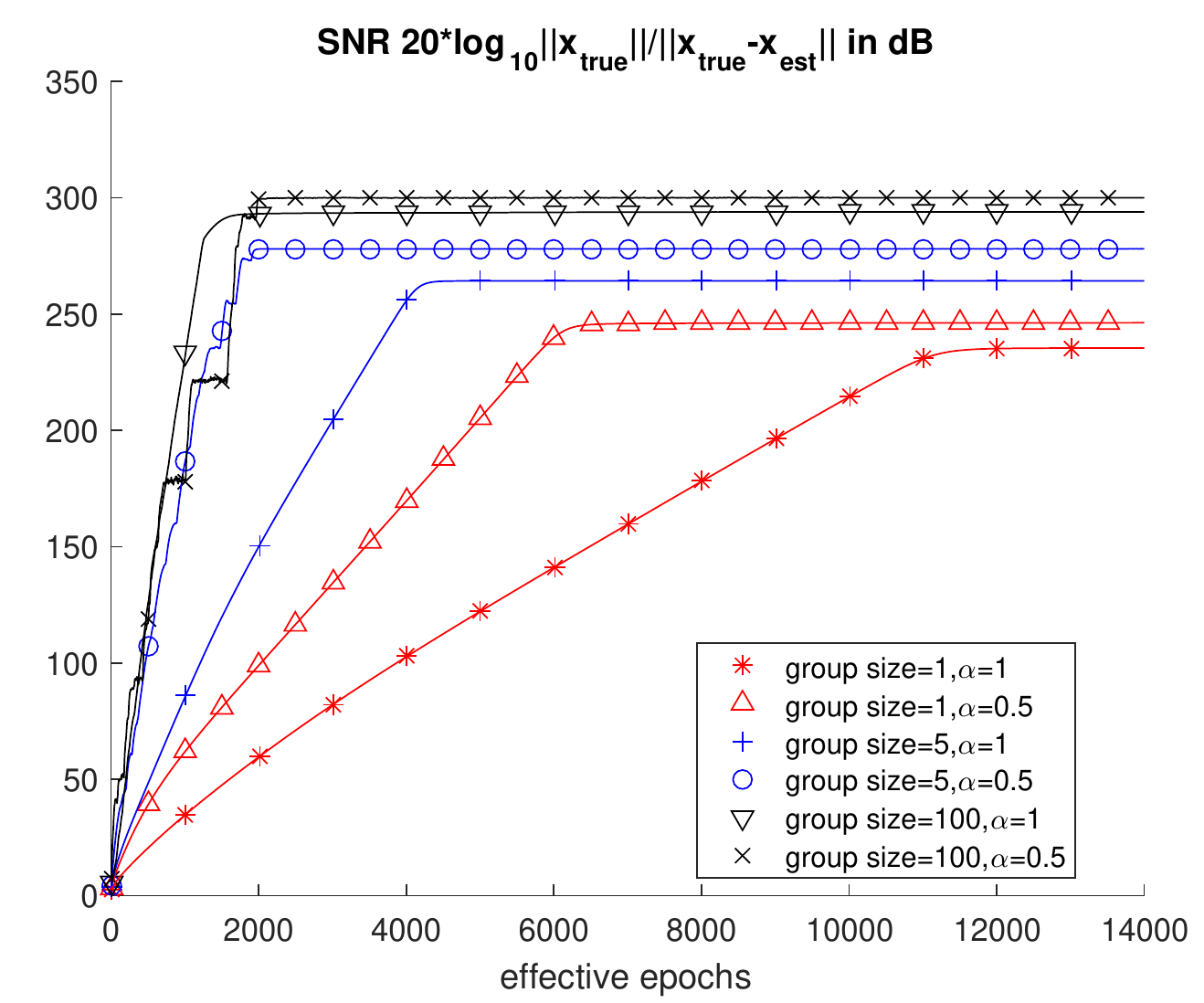}
\caption{ Different group sizes lead to different precision levels and different convergence rates. The parameter $b$ is tuned to ensure GCSGD with different group sizes does converge. From group size 1 to group size 100, the corresponding $b$ is 100,25,2 respectively. Some \emph{flat areas} in group size=100 appear and the reason will be explained later. The observation gap has the similar trend and is not presented here due to the limitation of paper length.}
\label{f5}
\end{figure}

\begin{figure}[!]
\centering     
\subfloat[SNR=3.44dB]{\label{simurea}\includegraphics[width=1.1in,height=1.1in]{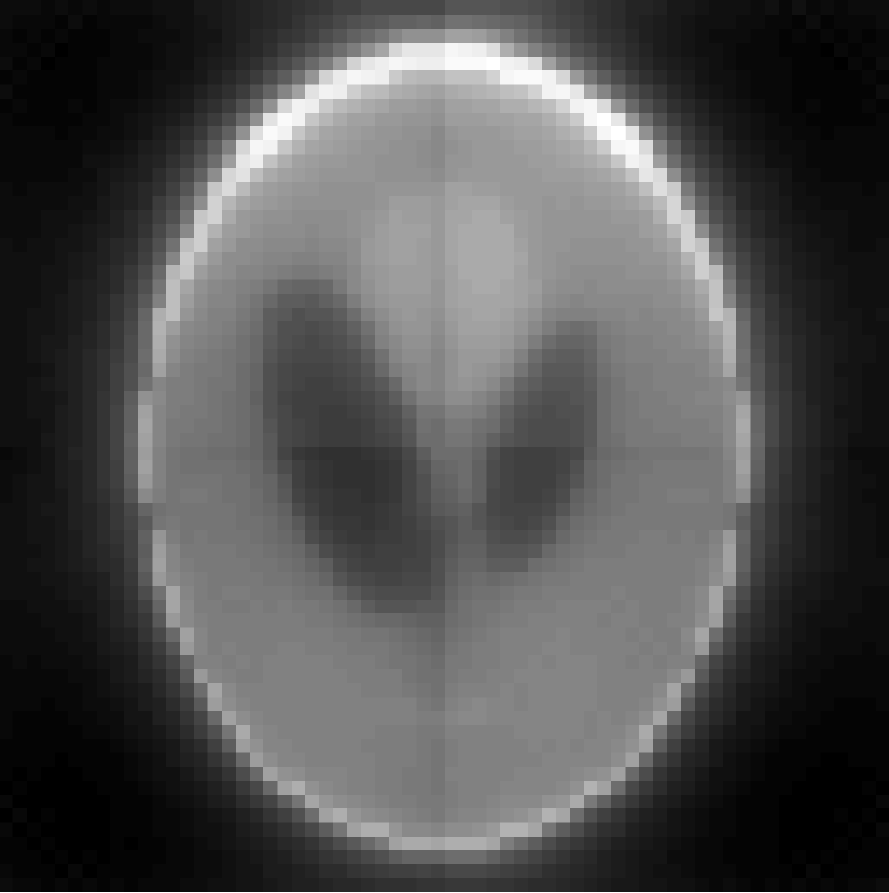}}
\subfloat[SNR=6.03dB]{\label{simureb}\includegraphics[width=1.1in,height=1.1in]{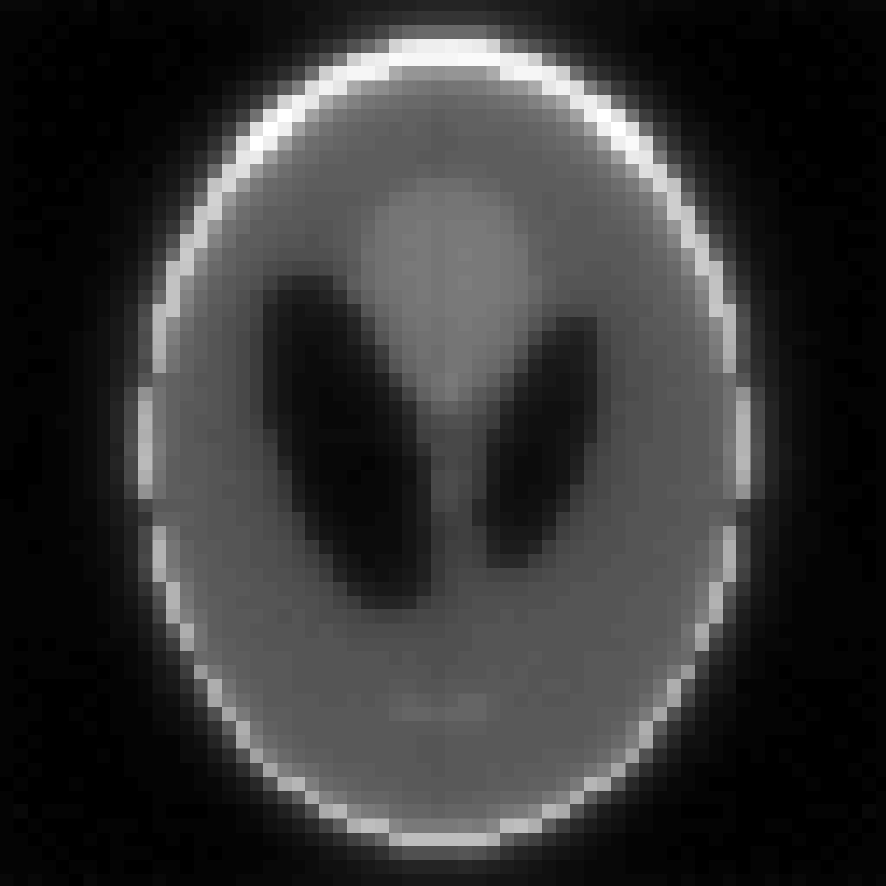}}
\subfloat[SNR=7.75dB]{\label{simurec}\includegraphics[width=1.1in,height=1.1in]{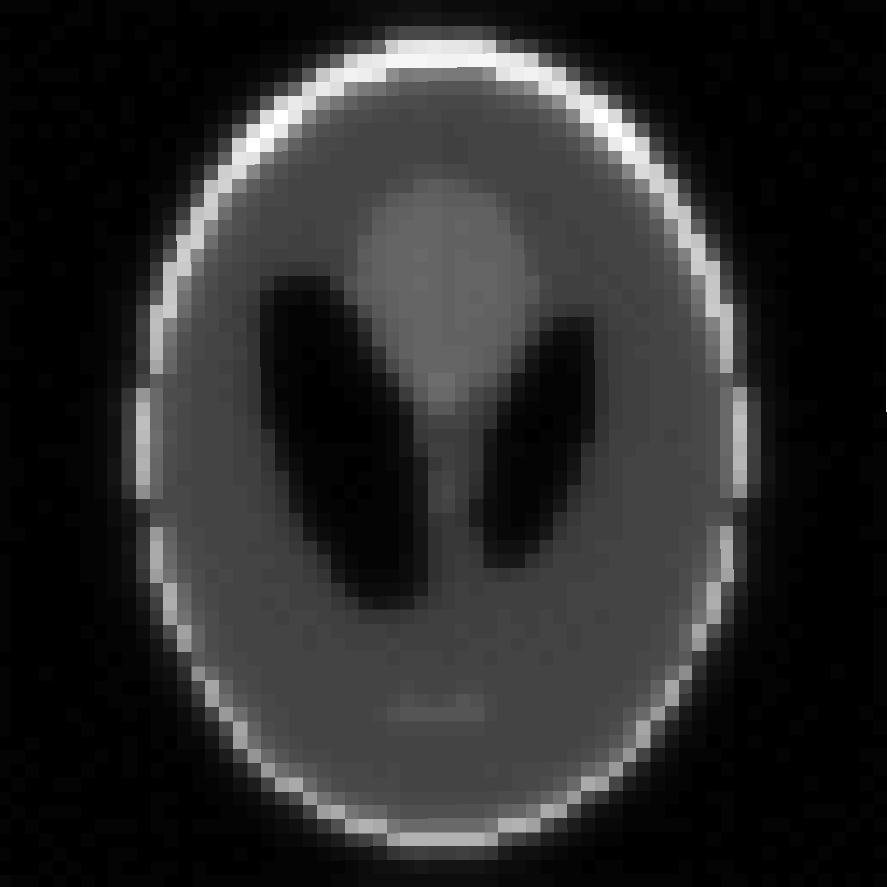}}
\\
\subfloat[SNR=5.43dB]{\label{simured}\includegraphics[width=1.1in,height=1.1in]{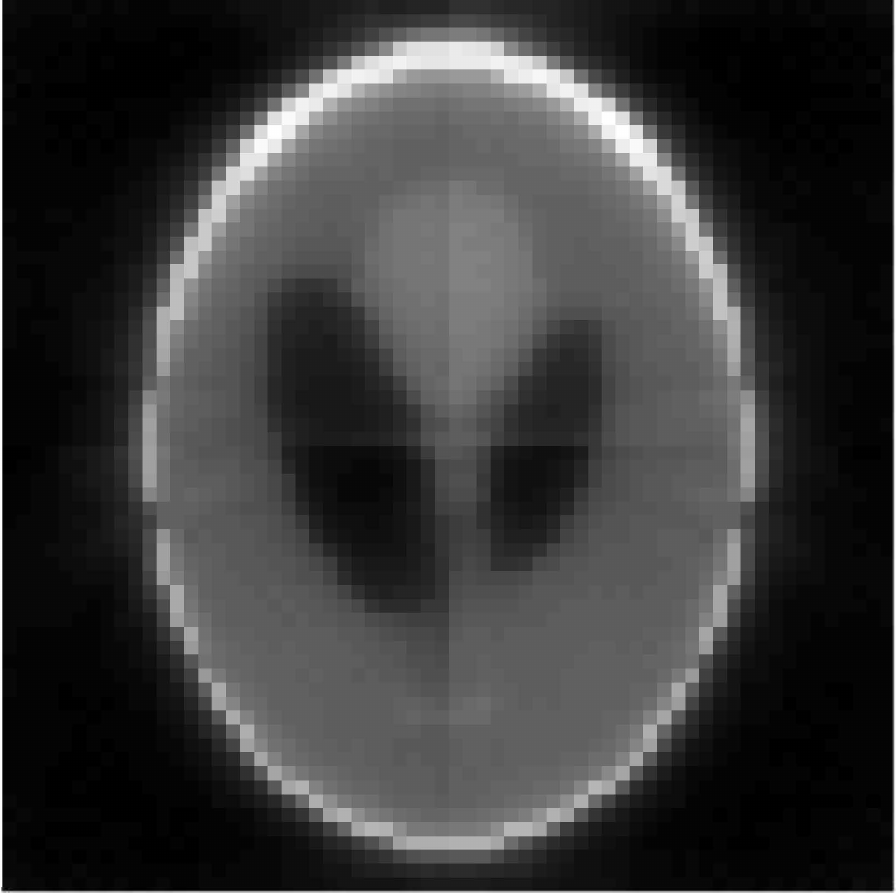}} 
\subfloat[SNR=11.42dB]{\label{simuree}\includegraphics[width=1.1in,height=1.1in]{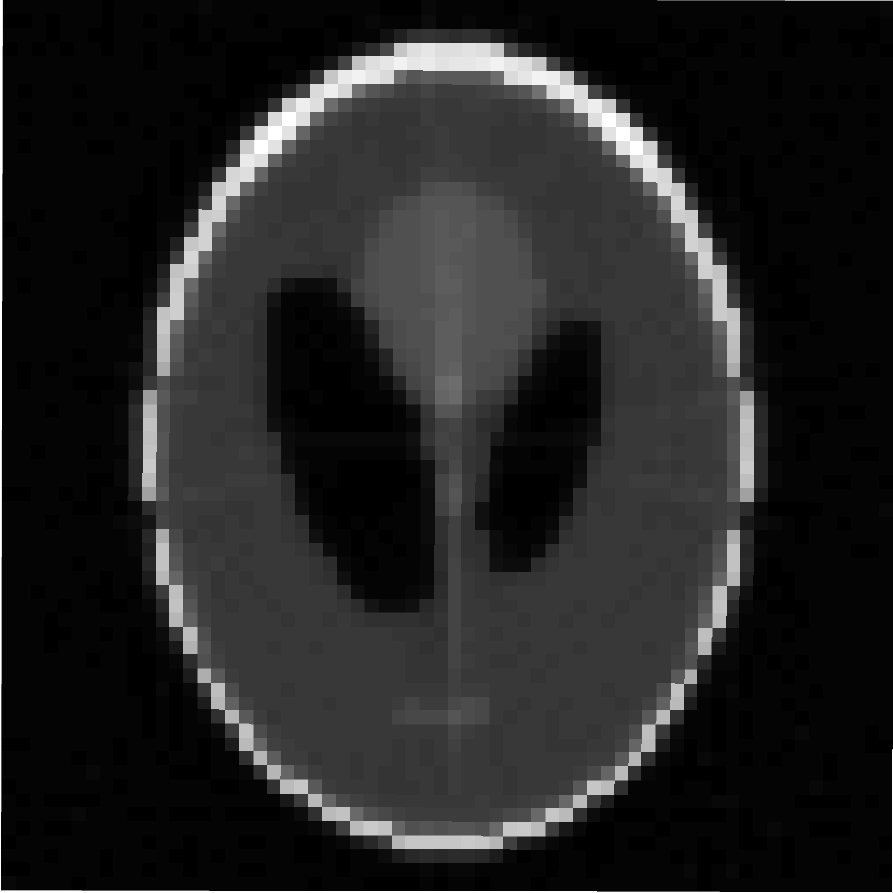}}
\subfloat[SNR=23.76dB]{\label{simuref}\includegraphics[width=1.1in,height=1.1in]{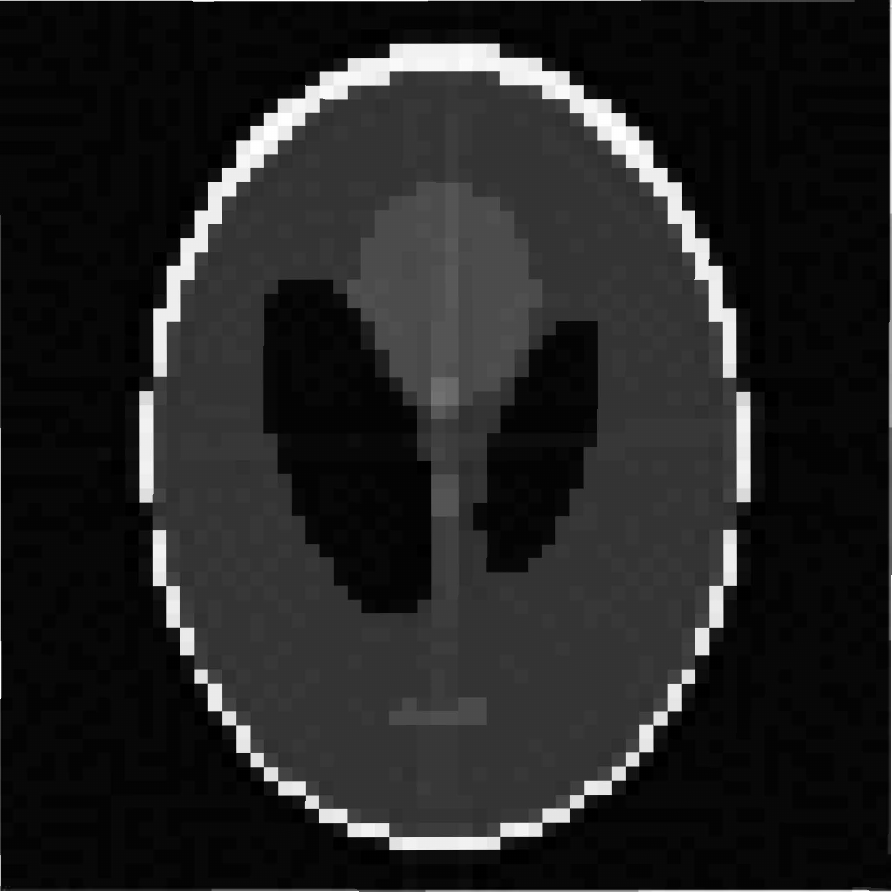}}
\caption{Simulation results after 20 effective epochs.(a) group size=1, $\a=1$, $b=100$; (b) group size=5, $\a=1$, $b=25$; (c) group size=100, $\a=1$, $b=2$; (d) group size=1, $\a=0.5$, $b=100$; (e) group size=5, $\a=0.5$, $b=25$;(f) group size=100, $\a=0.5$, $b=2$. When $\a$ is 0.5, the reconstructed images are blurred by artefacts. The reasons and the solutions will be discussed later.}
\label{simure}
\end{figure}

When we set $\a=1$, the importance sampling strategy cannot show any advantage since we always use all data. To demonstrate the difference, we set the group size to 1, $b=100$ and set $\a$ to 0.8,0.5 and 0.2 respectively. Fig.\ref{f15} shows that setting $\a$ smaller than 1 increases the convergence rate of the CSGD. Furthermore, different values of $\a$ always lead to the same precision. 
Considering that the smaller $\a$ is, the more actual epochs are required to achieve the same number of effective epochs and that MATLAB is not efficient in performing loops, in following simulations, we chose a moderate parameter $\a=0.5$ to test other qualities of CSGD and GCSGD. For example, we test the influence of group size of the GCSGD. Simulation results are shown in Fig.\ref{f5}. 
It can be seen that the GCSGD with importance sampling strategy all converges to a high precision level regard less of the group size. Generally speaking, setting the group size as large as possible helps to increase the initial convergence rate. This is reasonable since more row information is used for each epoch when the group size is enlarged, making the iteration more likely to move towards the global \emph{downhill} direction. Furthermore we empirically observed that a larger group size requires smaller parameter $b$ to maintain convergence. This is because when the group size increases, the ratio $P_S/P_T$ also increases and thus a smaller $b$ is required, otherwise it is easy for the step to overshoot, which makes the algorithm fail to converge. 
In practice, however, for parallel computing architectures, we suggest that the group size should be designed depending on the storage ability of each computing node. To show differences when using different group sizes and different $\a$s, some reconstruction results are presented in Fig.\ref{simure}.

\subsection{Different partitions}
\begin{figure}
\centering     
\subfloat[group size is 1, the $b$ is set as 100,200,400,800 respectively for detector areas being 2 to 16. ]{\includegraphics[width=3.5in]{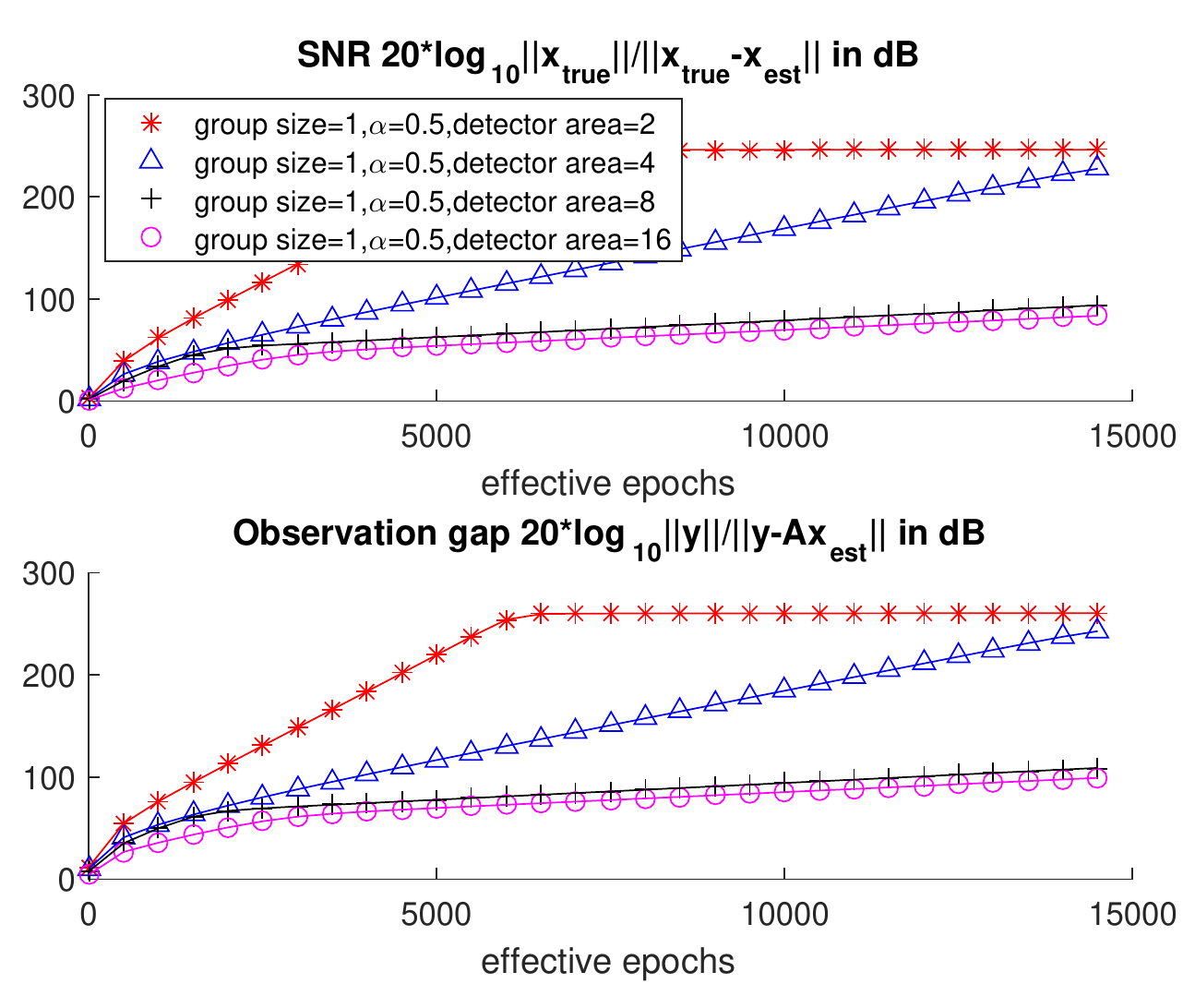}}

\subfloat[group size is 100, the $b$ is set as 2,4,8,16 respectively for detector areas being 2 to 16. ]
{\includegraphics[width=3.5in]{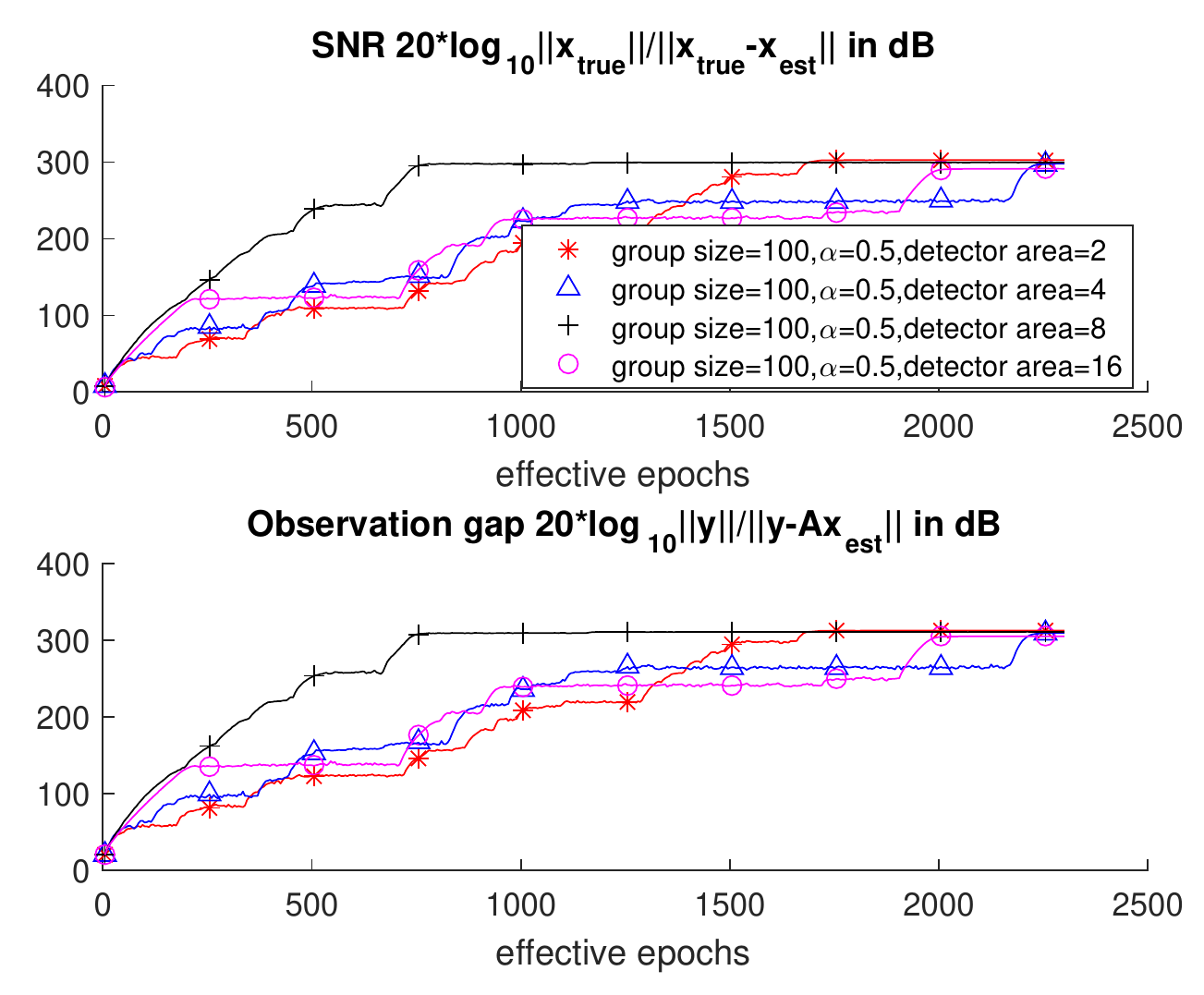}}
\caption{The convergence results for different detector partition numbers. A larger group size is of help to increase the convergence rate regardless of the appearance of flat areas.}
\label{f7}
\end{figure}

The above simulations show the effectiveness of the CSGD and GCSGD method with importance sampling. In following simulations, we explored the influence of different partition methods. For example, the reconstructed image was still divided into 4 square blocks and the detector area was partitioned into 2,4,8 and 16 sets respectively. We compared the difference between group size being 1 and 100 to further verify that a larger group helps to increase the convergence rate. The simulation results are shown in Fig.\ref{f7}.

When the group size is only one, we can see that dividing the detector into 8 and 16 areas leads to rather slow convergence rates. It suggests that when the group size is small, the detector cannot be divided into more sub-areas. For a group size of 100, however, the algorithm allows for a devision of the detector into more areas and the final precision level is not affected. In the initial iterations, we can see that different detector partitions have nearly the same convergence rate in the GCSGD method. When the precision level is high, the \emph{flat areas} appear again and the reason will be explained later. In fact, the \emph{flat area} is of less interest to us since it only appears when the reconstructed image already has a very high precision level and never appears for the initial iteration stage. As a result, this simulation suggests that within a range, reducing the row information (partition the detector area from 2 to 4 or even bigger when group size is fixed) does not influence the initial convergence rate when using GCSGD, especially when the group size is large. It is straightforward to see that the calculation amount of $A_I^J$ for a selected sub-volume and sub-detector area decreases when the number of partitions on the detector increases. As a result, when addressing the partition strategy for large scale tomography problems, a more flexible partition method is allowed for large group size situation. If the big group sizes poses a heavy storage burden for work nodes, a finer partition method would be of help to reduce the computation amount for $\mathbf{A}_I^J$. In other words, although dividing the detector area into more sub-areas cannot provide a higher precision level, it helps to reduce the computation load when only a few iterations are performed. 

\begin{figure}
\centering     
\includegraphics[width=3.5in]{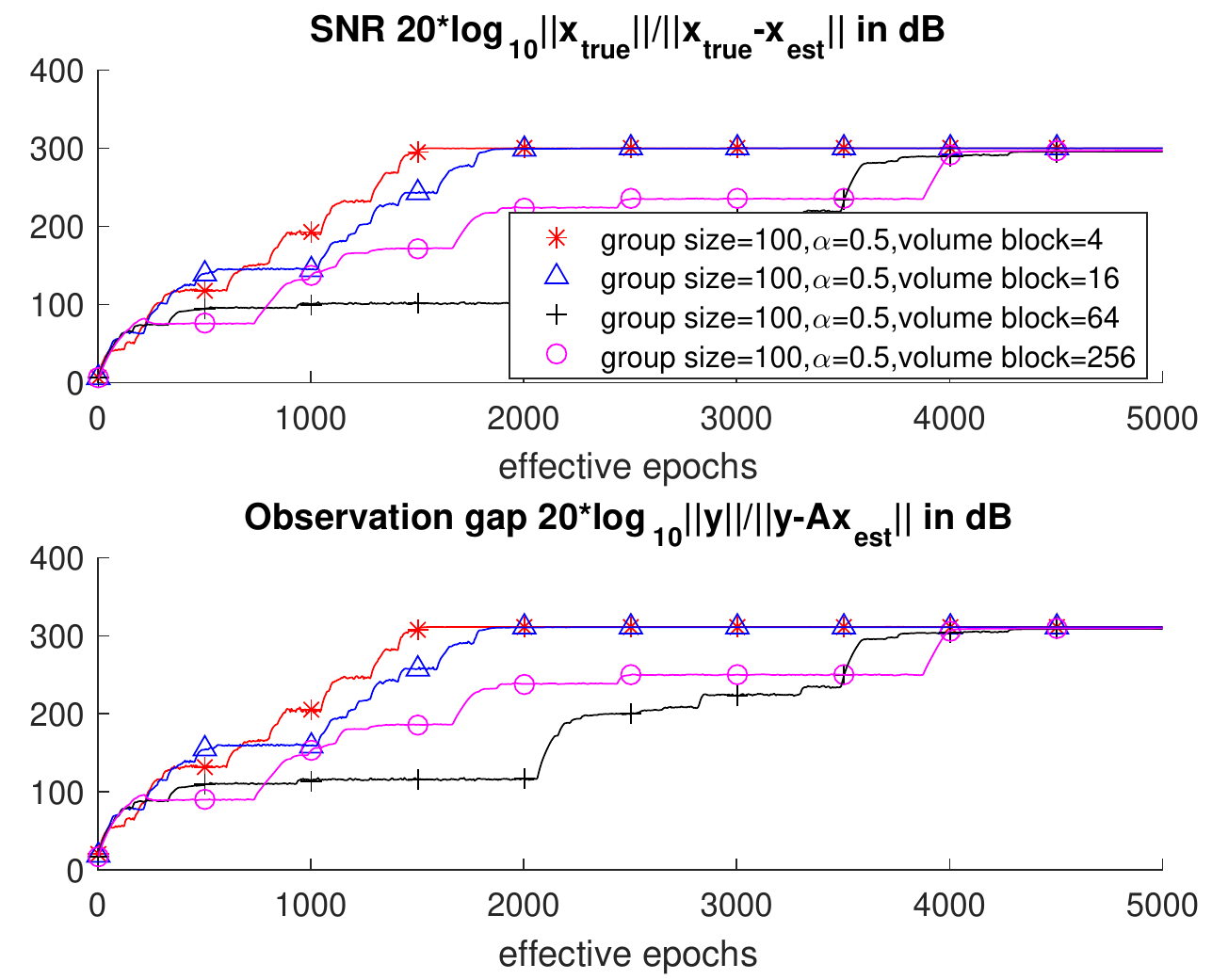}
\caption{When the image is divided into different volume slices, and all volume blocks are selected for each epoch, the initial convergence rates and the final precision levels are similar to each other regardless of flat areas.}
\label{f16}
\end{figure}

Another variable is the number of volume blocks. To investigate this, we partitioned the 2D image into 4,16,64,256 areas respectively (divide each dimension into 2,4,8,16 parts evenly). 
The detector area was always partitioned into 2 sub-areas, as shown in Fig.\ref{f14}. The group size was set to 100 and the corresponding $b$ was 2,1,0.5,0.25 respectively to ensure convergence. The simulation results are shown in Fig.\ref{f16}. It can be seen that all partition methods show similar final precision level, and in the initial iterations, the differences between different volume block numbers are also negligible, demonstrating the good scalability of this algorithm. Since a finer partition allows more computation nodes to be used and the computation load of each node is reduced, the similar convergence in the initial iterations is useful when addressing large scale data where only a few iterations are performed.

\begin{figure}
\centering     
\subfloat[When the image is partitioned into 16 square blocks, $b=1$ for all simulations]{\label{f10a}\includegraphics[width=3.5in]{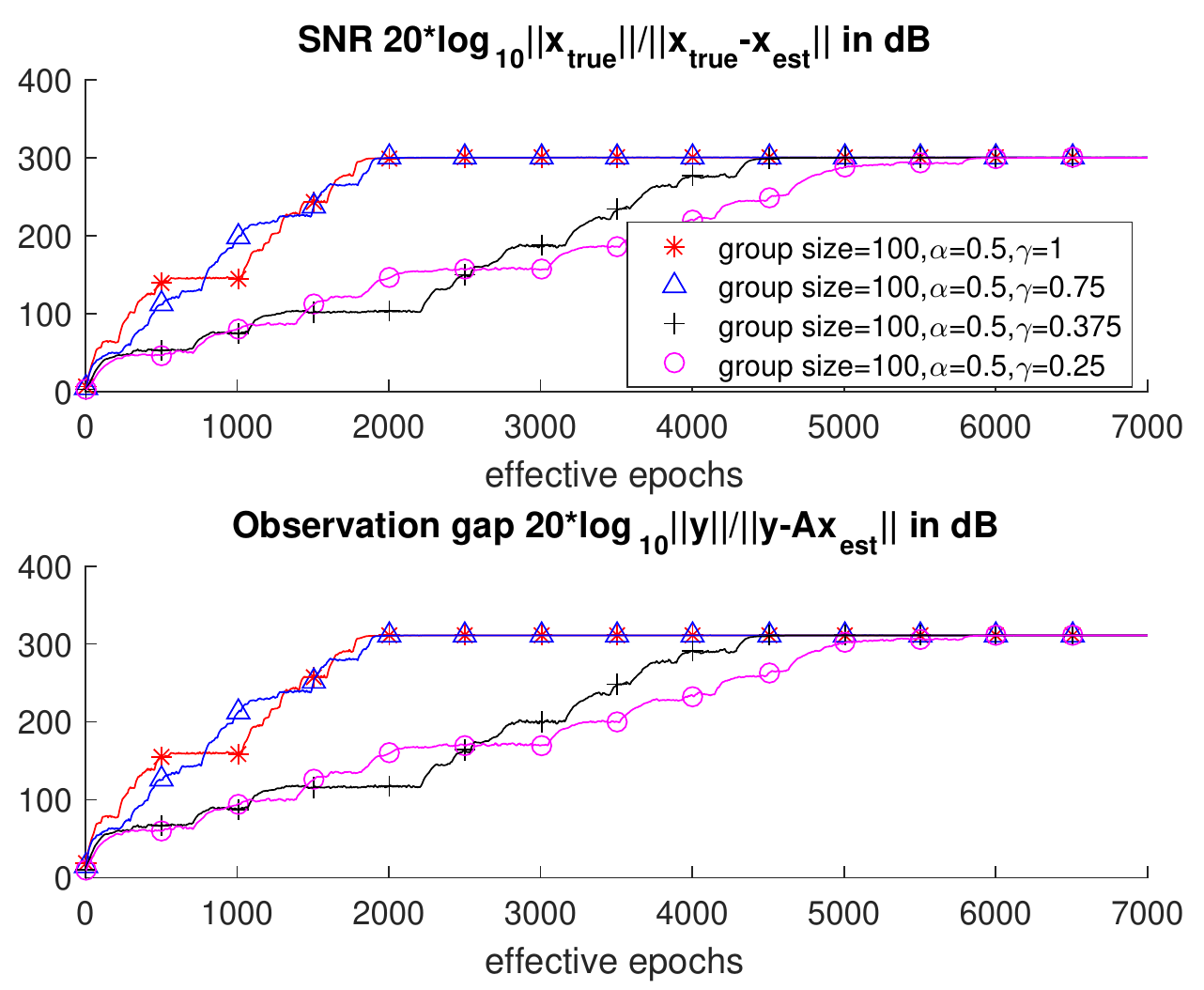}}

\subfloat[When the image is partitioned into 64 square blocks,$b=0.5$ for all simulations]{\label{f10b}\includegraphics[width=3.5in]{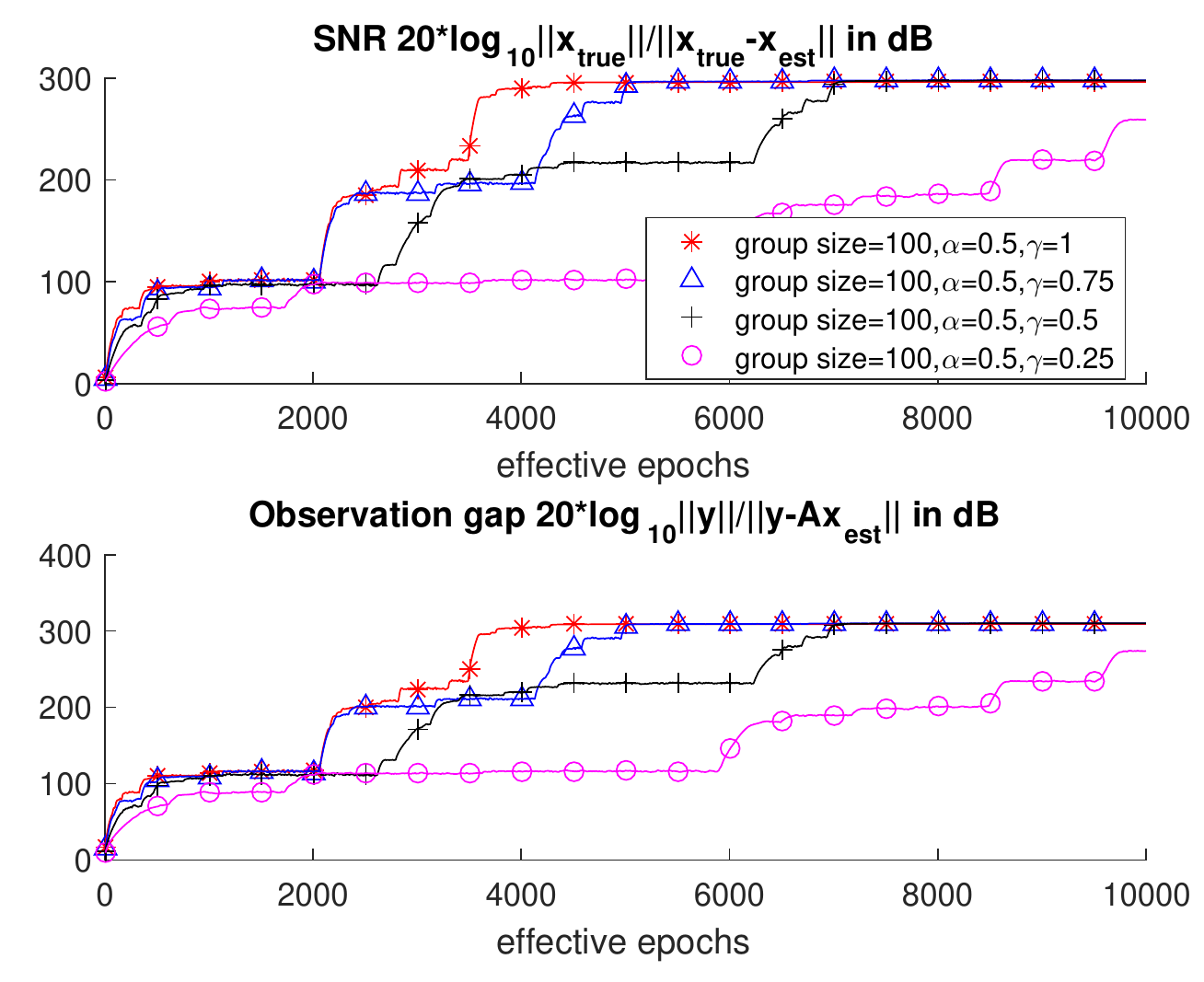}}
\caption{When randomly select $\gamma*100\%$ image blocks to get involved in iteration for each epoch. Despite of the flat areas, the final precision levels between different $\gamma$ are the same. The initial convergence rates are similar when the $\gamma$s are large than 0.5. When $\gamma$ is 0.25, the convergence rate slows down.}
\label{f9}
\end{figure}

We also explored the situation when not all of  $J$ were selected. The group size was set to 100 and the number of detector sub-areas was set to 2 by default. We set $\gamma$ as different values when the image were partitioned into 16 and 64 sub-areas, simulation results are shown in Fig.\ref{f9}. 

It can be seen that GCSGD method converges for every $\gamma$, and their final precision level is similar. Also, as long as the $\gamma$ is not too small ($\gamma=0.2$), then the initial convergence rates for different $\gamma$ are almost identical. This suggests that randomly leaving a small portion of the image blocks not to be updated does not influence the initial convergence rate and final precision level. This property will be useful when we perform the algorithm in parallel computing architecture where the algorithm maybe run asynchronously.  

\subsection{Discussions on importance sampling strategy}

In this part we compared the importance sampling and random sampling strategy based on the previous simulation geometry. Since the convergence property has been verified in previous simulations, in this part, we were only interested in the initial convergence rate and thus we only perform the algorithm (with importance sampling or random sampling strategy) for 50 effective epochs but repeat it 10 times for the same data and then calculate the average of SNR and observation gap.

The simulations not only show the drawbacks and advantages of the importance sampling strategy, but also explain why the \emph{flat area} appears when the precision level is high and group size is large, and why using importance sampling can lead to images artefacts.

\begin{figure}
\centering     
\includegraphics[width=3.5in]{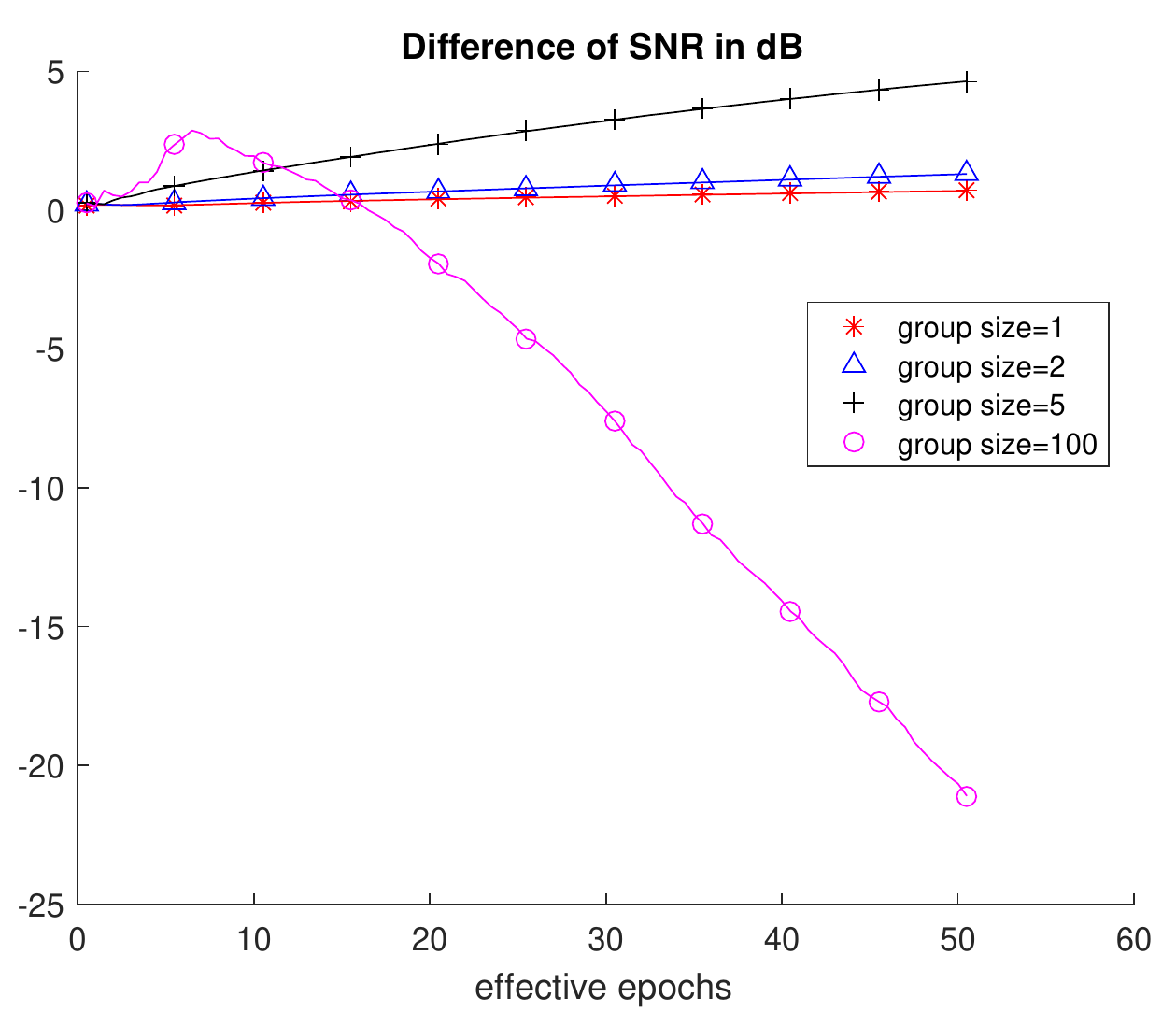}
\caption{The importance sampling strategy always outperforms random sampling when group size is 1, 2 and 5. When the group size is 100, the importance sampling strategy only works at the early stage, and then the advantages over random sampling disappears. The observation gap has the similar trend and is not presented here due to the limitation of paper length.}
\label{ComImRanGs1and2and5}
\end{figure}

\begin{figure}
\centering     
\includegraphics[width=3.5in]{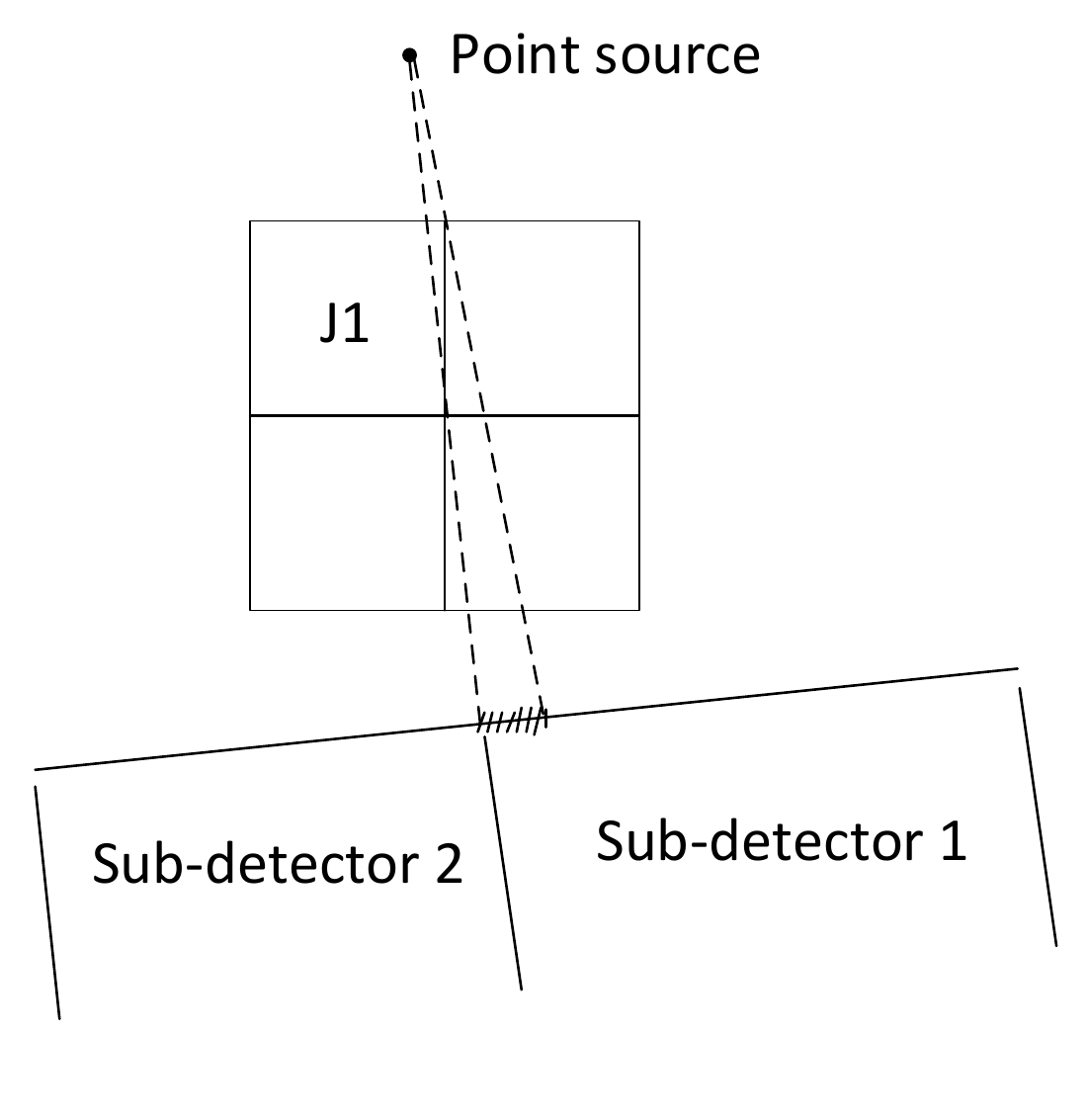}
\caption{The difference of the projection area for one projection view can be huge for some locations. Here $I_1$ means the row blocks for sub-detector 1 in the current projection view and $I_2$ has similar definitions for sub-detector 2.}
\label{Geometry}
\end{figure}

We first verified that the importance sampling strategy indeed helps to increase the convergence rate compared with random sampling. However, it only happens when the group size is small. In simulation, we applied two sampling strategies to sample the sub-detector areas and calculate the averaged SNR for each iteration. Two dataset $SNR_{Imp}$ and $SNR_{Ran}$ were obtained, where the former one was the SNR through importance sampling and the latter one was obtained through random sampling. The difference $SNR_{Imp}-SNR_{Ran}$ is shown in the Fig.\ref{ComImRanGs1and2and5}. 
The reason why importance sampling cannot outpeform random one when group size is 100 is due to the non-uniform update on $\z_I^j$. For each volume slice, the selection criteria for choosing row blocks (or sub-detector areas from different projection views) is based on the projection area, which reflect the sub-matrix's density. The difference between projection areas can sometimes be huge. For example, as shown in Fig.\ref{Geometry}, for volume slice $J_1$, the projection area on the sub-detector 2 is far larger than that on the sub-detector 1. That means when updating $J_1$ and selecting sub-detector areas, or row blocks, it is more likely that sub-detector 2 is chosen. This property makes the update of $\z_{I_2}^{J_1}=\A_{I_2}^{J_1}\x_{J_1}$ very frequent and the update on $\z_{I_1}^{J_1}=\A_{I_1}^{J_1}\x_{J_1}$ rather infrequent. For residue terms $\r_{I_1}$ and $\r_{I_2}$, the contribution from $\z_{I_1}^{J_1}$ and $\z_{I_2}^{J_1}$ is different, especially in the initial iterations. Since the density of $\A_{I_2}^{J_1}$ is higher than $\A_{I_1}^{J_1}$, this means the influence of $\z_{I_2}^{J_1}$ on $\r_{I_2}$ is more significant than $\z_{I_1}^{J_1}$ on $\r_{I_1}$. At the initial iterations, the frequent updates of $\z_{I_2}^{J_1}$ make $r_{I_2}$ decrease efficiently whereas the infrequent updated of $\z_{I_1}^{j_1}$ has little influence on the update of $\r_{I_1}$. (It should be remembered that at the same time $\r_{I_1}$ is  also updated by computations with the other volume slices.) This property helps the importance sampling strategy to obtain a fast initial convergence rate but when the precision level is already high, the update on $\r_{I_1}$ then needs the contribution from $\z_{I_1}^{J_1}$ and unless this term is updated frequently, then $\r_{I_1}$ cannot be further reduced. This explains why at the initial stage the importance sampling strategy is faster than random sampling and when the iteration goes on, there are some \emph{flat areas} in Fig.\ref{f7}, Fig.\ref{f16}, Fig.\ref{f9}. The reason for the artefacts in the reconstructed images is similar: the shaded area in  Fig.\ref{Geometry} is rarely used to update  $J_1$, thus allowing the reconstruction error of the inner margin of $J_1$,(the same holds in the other volume slices) a little higher than the other parts. The reason why the importance sampling strategy works and the \emph{flat areas} disappear when the group size is small (for example, group size is set as 1) is because that in this situation, the total convergence rate is slow and when the precision level reaches a high level, then the update on $\z_{I_1}^{J_1}$ is already frequent enough to allow an effective change on $r_{I_1}$.   

\begin{figure}
\centering     
\includegraphics[width=3.5in]{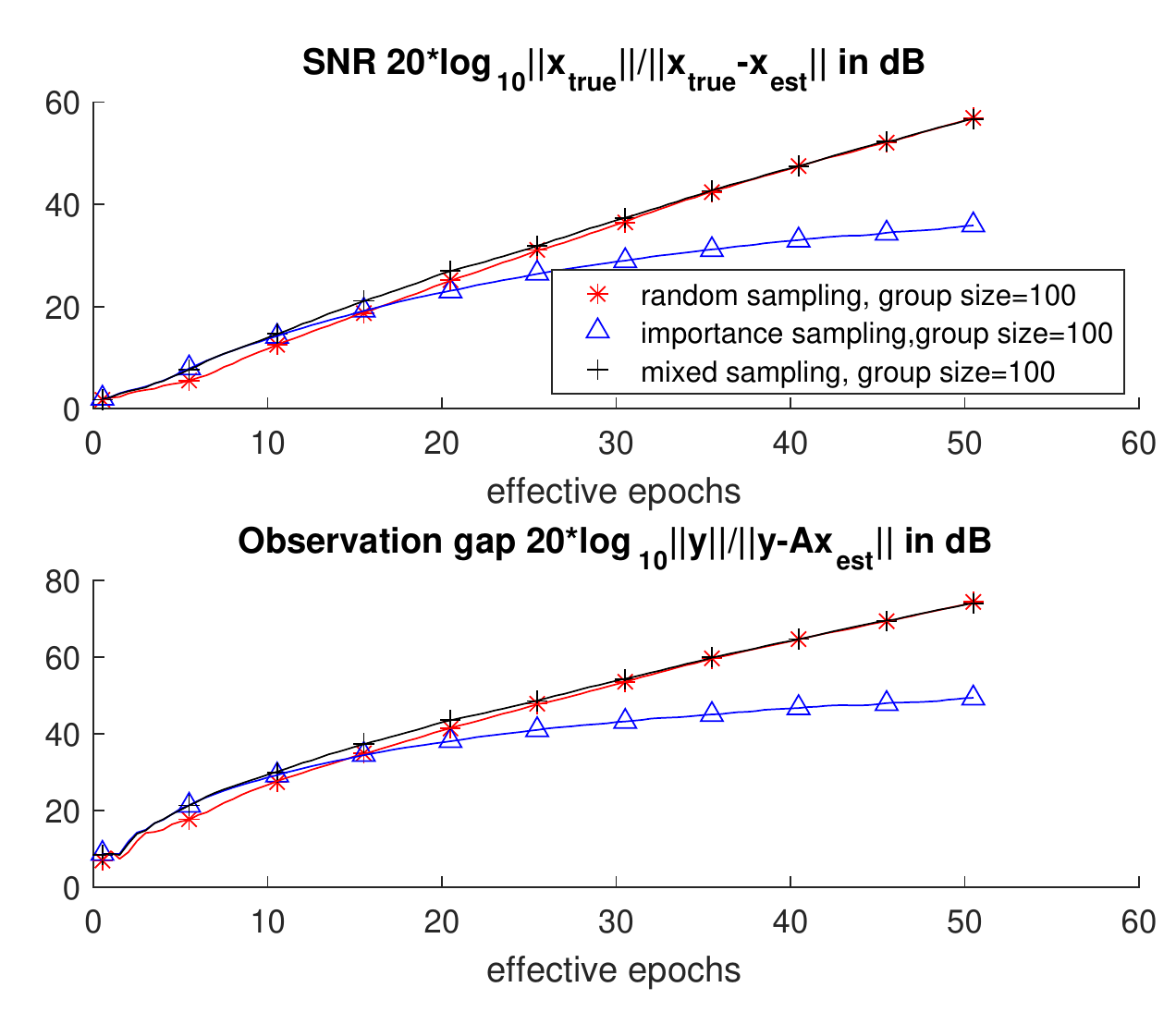}
\caption{Comparisons for different sampling strategies. The $\theta$ in mixed sampling method starts from 0 and the increment step length is 1/40, it means that after 40 epochs the probability distributions on different sub-detector areas becomes the same within one projection. The mixed sampling method maintains the same convergence rate at the initial stage as importance sampling strategy does, and this increment trend is maintained along with the iteration going on.}
\label{ComImRanMixGS100}
\end{figure}

We proposed some methods to overcome the disadvantages of importance sampling. One method would be to allow the algorithm to make a choice before sampling the row blocks, deciding whether to use importance or random sampling in the current iteration. This choice can be done with a specific probability. Another method would define a threshold in the probability distribution, enforcing it to be above the threshold. However, simulation results show that these two methods are not very effective in terms of improving the performance of importance sampling strategy. The third method, we call a mixed sampling strategy, is to gradually change the probability distribution along with the iterations. By reducing the gap between different probability values gradually, we experimentally showed that this method can maintain the fast initial convergence due to the importance sampling and eliminate the artefacts in the reconstructed image and the \emph{flat area} in the convergence line. Specifically, in Fig.\ref{Geometry}, we defined the projection area of $J_1$ on sub-detector area 1 as $P_1$ and that on sub-detector area 2 as $P_2$. In the previous importance sampling strategy, the probability for choosing sub-detector area 1 is $P_1/P_T$, where $P_T$ is the total projection area of sub-volume $J_1$. Similarly, the probability for choosing sub-detector 2 is $P_2/P_T$. The mixed sampling method gradually fills the gap between $P_1/P_T$ and $P_2/P_T$. The probabilities to choose sub-detector 1 and 2 are set as $[P_1+\theta(P_{max}-P_1)]/P_T$ and $[P_2+\theta(P_{max}-P_2)]/P_T$ respectively, where $P_{max}$ is the largest value of $P_1$ and $P_2$. The parameter $\theta$ is used to reduce the difference between two possibilities by gradually increasing it from 0 to 1. We make a comparison with importance sampling, random sampling and the mixed sampling method when the group size is 100 and $\a=0.5$. The convergence rate is shown in Fig.\ref{ComImRanMixGS100}.

\begin{figure}[!]
\centering     
\subfloat[SNR=4.90dB]{\label{simurea}\includegraphics[width=1.1in,height=1.1in]{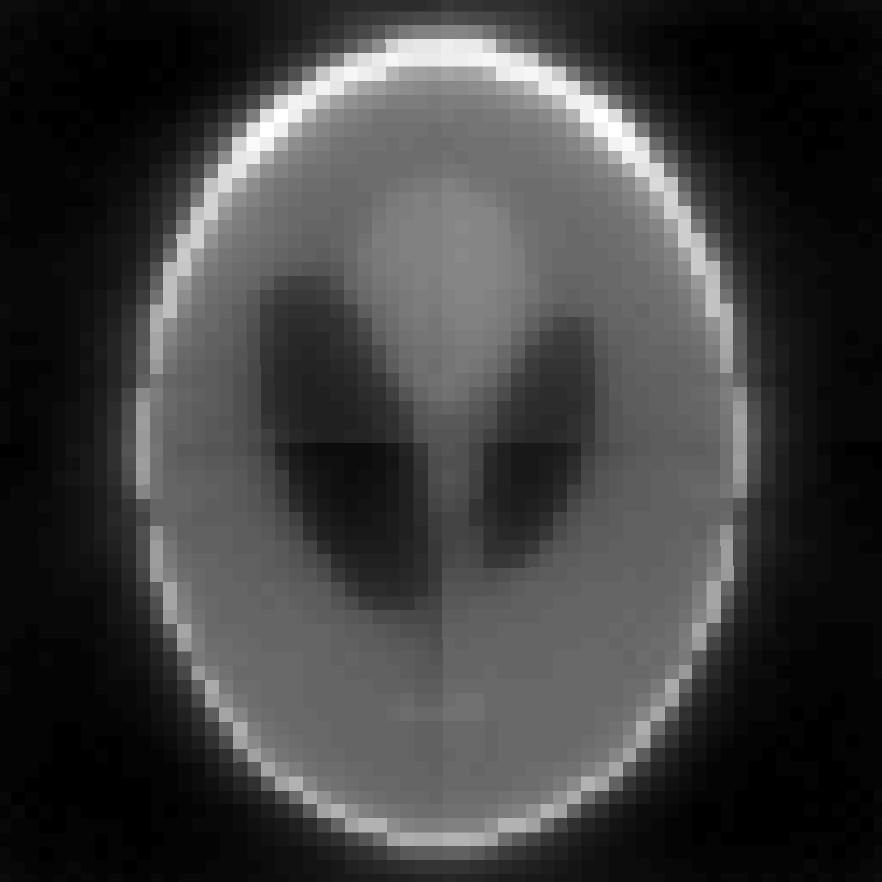}}
\subfloat[SNR=10.12dB]{\label{simureb}\includegraphics[width=1.1in,height=1.1in]{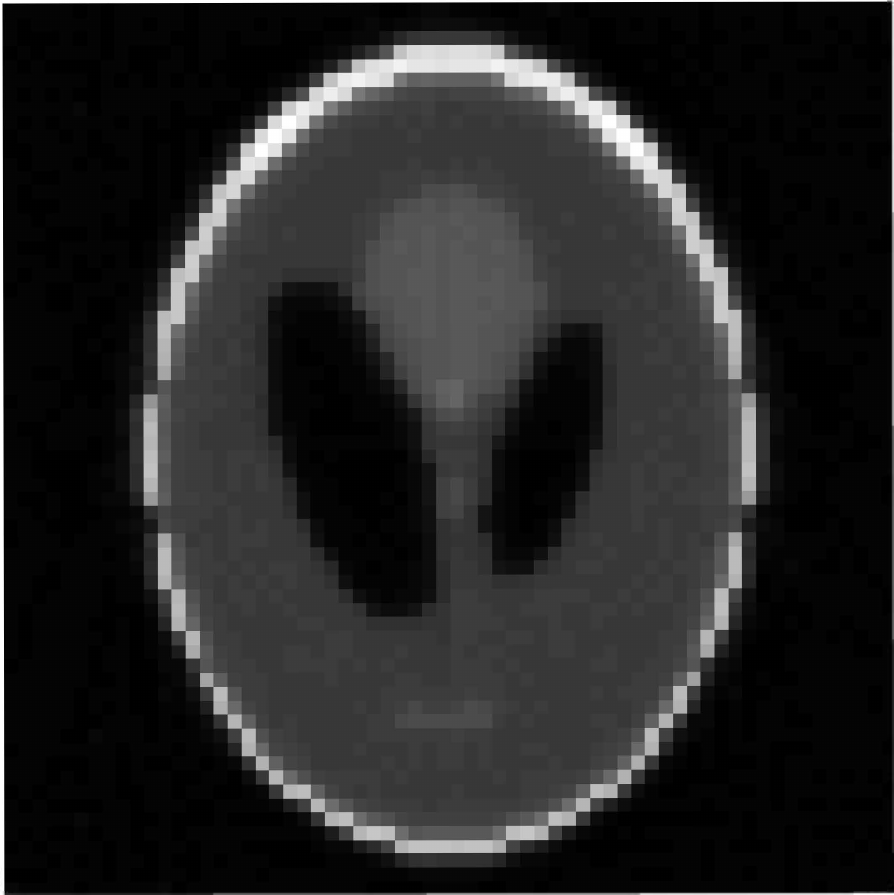}}
\subfloat[SNR=26.44dB]{\label{simurec}\includegraphics[width=1.1in,height=1.1in]{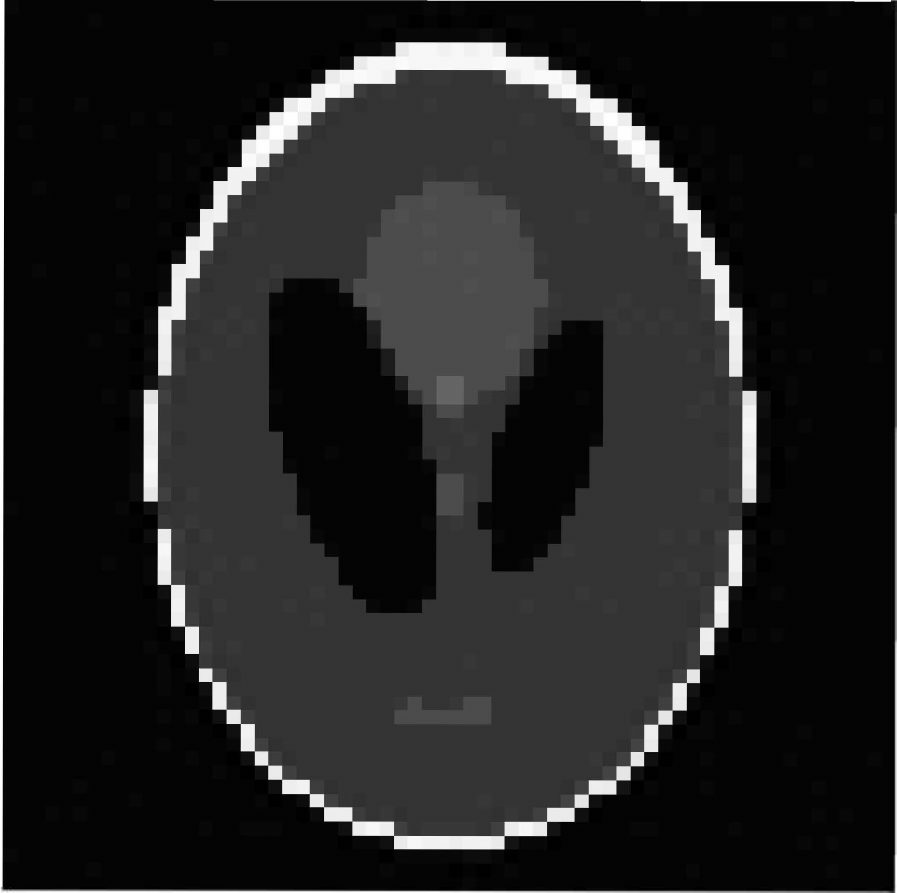}}
\caption{Simulation results after 20 effective epochs when use mixed sampling strategy.(a) group size=1, $\a=0.5$, $b=100$; (b) group size=5, $\a=0.5$, $b=25$; (c) group size=100, $\a=0.5$, $b=2$. Compared with Fig.\ref{simure}, the reconstructed images for group size 5 and 100 are free from the influence of artefacts while keeping the similar SNRs.}
\label{simuremix}
\end{figure}

We also present the reconstructed image after 20 effective epochs for group size being 1,5,and 100 respectively. The simulation condition is the same as Fig.\ref{simure}. For simplicity, we only simulated the $\a=0.5$ situations. The simulation results are shown in Fig.\ref{simuremix}.

\begin{figure}
\centering     
\includegraphics[width=3.5in]{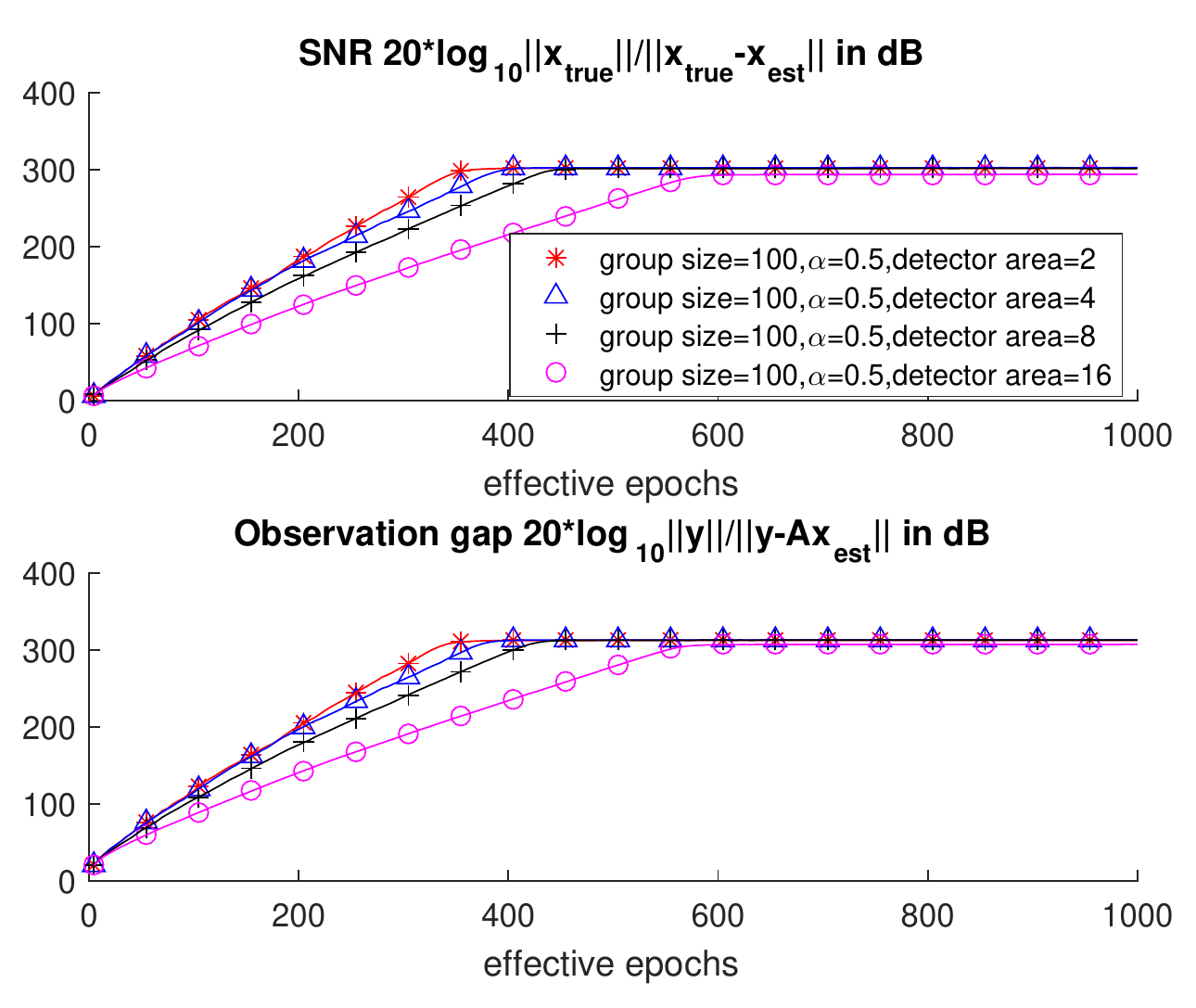}
\caption{Using the mix sampling strategy eliminate the \emph{flat area}, thus shorten the iteration numbers needed to achieve the precision limit.}
\label{ComDiffDetec}
\end{figure}

The mixed sampling method can also eliminate the \emph{flat area} in the convergence trend. For example, we here present the simulation results of different detector partition methods, as shown in Fig.\ref{ComDiffDetec}. Compared with Fig.\ref{f7}, the convergence trend is more smooth and it requires less iterations to achieve the precision limit.

\subsection{3D simulation}
\begin{figure}
\centering     
\subfloat[]{\label{f11a}\includegraphics[width=1.7in]{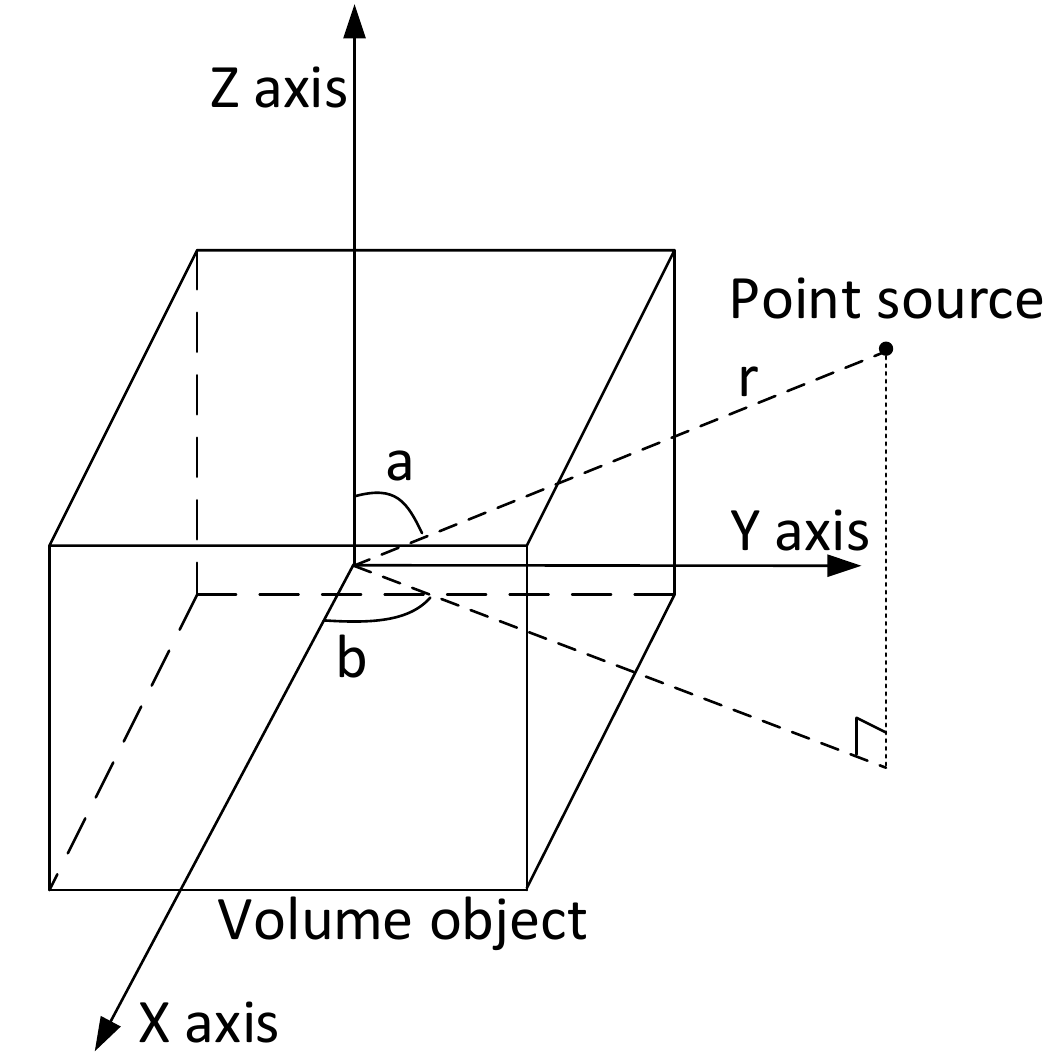}}
\subfloat[]{\label{f11b}\includegraphics[width=1.7in]{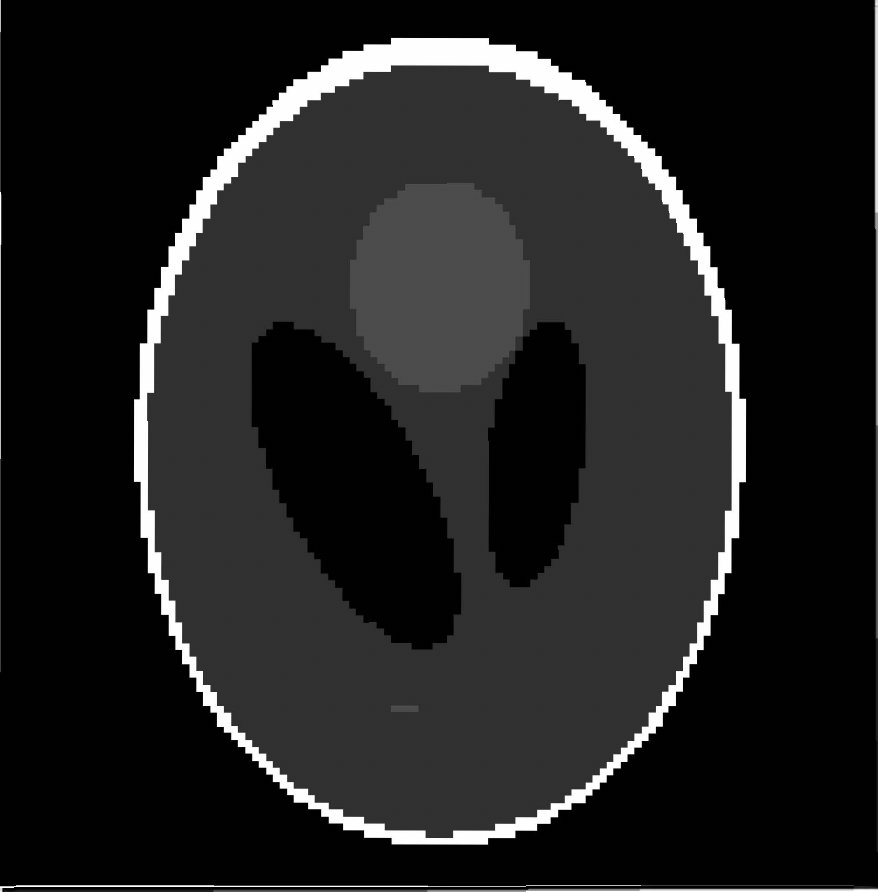}}
\caption{Basic 3D simulation settings. (a) shows that the location of the point source is $x=r\sin a \cos b$,$y=r\sin a \sin b$,$z=r\cos a$, where $a$ and $b$ are randomly changed. The connection line of point source and the centre of the detector plane is always perpendicular to the detector plane. (b) is a slice of the 3D volume.}
\label{f11}
\end{figure}

To demonstrate the performance in a 3D setting, we defined a 3D phantom of 128*128*128 voxels describing a 32*32*32 volume. The CCD detector plane was a square with side length 101 whose detector spacing was 0.5. To demonstrate how our method can be used for nonstandard trajectories, we chose a randomised scanning path with 720 projection views, i.e. the point source location was randomly changed with a fixed radius $r$ of 66 but with a random rotation. The rotation center was placed at the centre of the 3D object. The detector was always perpendicular to the line connecting the geometric centre of the detector and the point source, and the vertical distance between point source and detector plane was 132. The illustration of the scanning geometry and a slice of the 3D volume is shown in Fig.\ref{f11}. This simulation was performed on a computer with 48 Intel Xeon CPU cores and 256GB RAMs.

\begin{figure}
\centering     
\subfloat[SNR=3.40dB]{\label{f12a}\includegraphics[width=1.7in,height=1.7in]{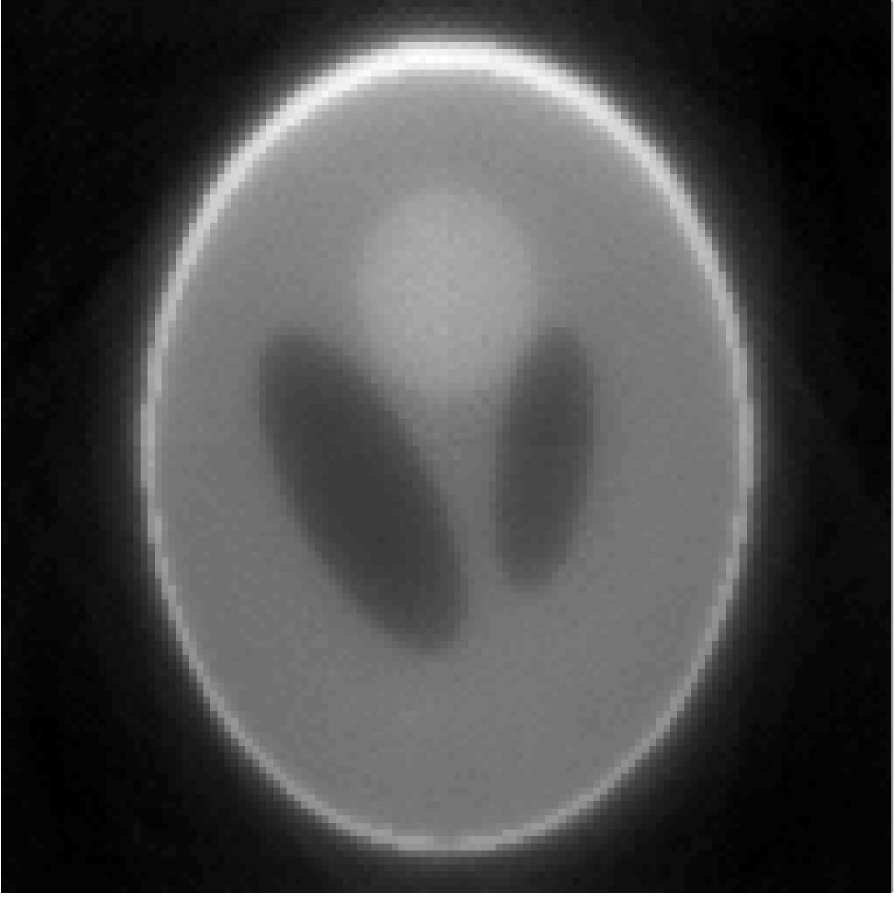}}
\subfloat[SNR=14.74dB]{\label{f12b}\includegraphics[width=1.7in,height=1.7in]{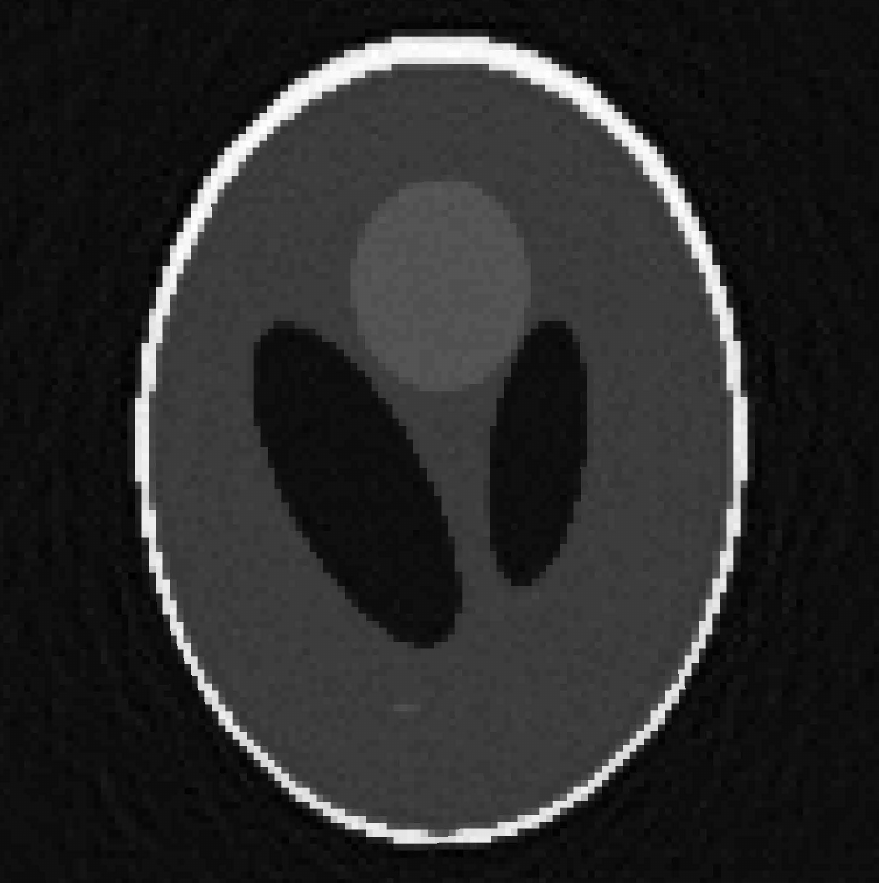}}\\
\subfloat[SNR=19.87dB]{\label{f12c}\includegraphics[width=1.7in,height=1.7in]{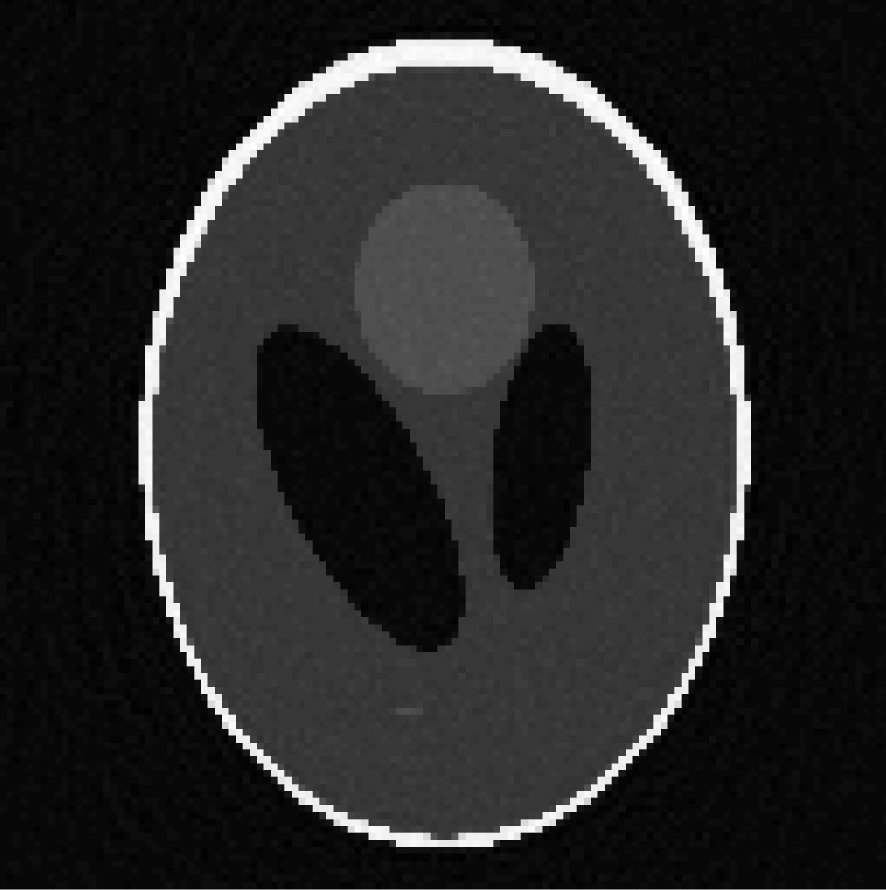}}
\subfloat[SNR=26.49dB]{\label{f12d}\includegraphics[width=1.7in,height=1.7in]{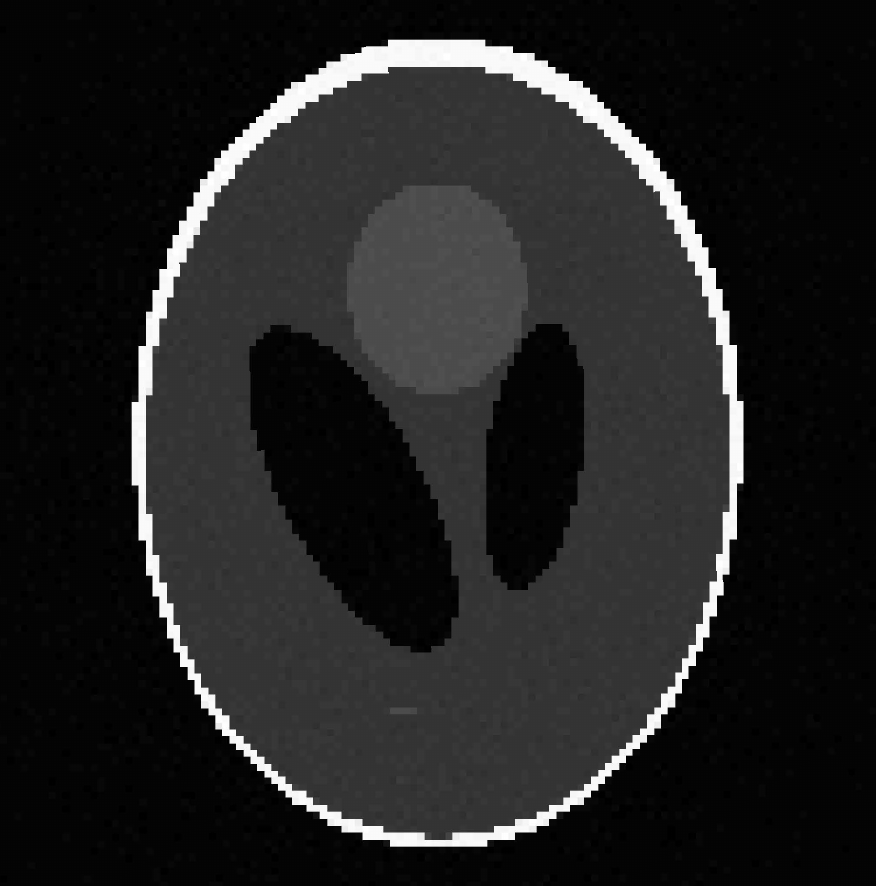}}

\caption{The selected slice of the reconstructed image after (a)5,(b)50,(c)100,(d)200 effective epochs.}
\label{f12}
\end{figure}

In this simulation, we did not divide the CCD area into sub-areas but simply treated each projection as a whole. We divided the volume into 2 sections for each dimension, thus we had 8 blocks in total. We set the group size to 90. $b$ in $\beta$ is set to 1, $\a=0.5$ and $\gamma=1$. Reconstruction results after different epochs are shown in Fig.\ref{f12}.

\section{Conclusion}
Here we propose a parallel algorithm CSGD that is suitable for linear inverse problems $\y=\A\x+\mathbf{b}$ and enables additional flexibility in the way we can partition data sets in  algebraic CT image reconstruction. As far as we know, CSGD is the first algorithm that combines row action methods and column action methods. Unlike other algorithms that demand the slice strategy of the reconstructed volume to be perpendicular to the rotation axial of the object to reduce data sharing between different computing nodes and a circular or helical scanning strategy, CSGD does not pose any restrictions on how to partition the reconstruction volume. It is thus applicable to generic scan trajectories. Furthermore, the CSGD method can be implemented using a server parameter parallel computing architecture. It thus has the potential to solve very large image reconstruction problems by using several computing nodes. The GCSGD method better utilises row information of the system matrix $\A$ and thus has faster convergence rates than CSGD. We have furthermore developed an importance sampling strategy. By non-uniformly selecting row blocks of $\A$ based on the projection area of the associated sub-volume, the sub-matrices of $\A$ with relatively higher density get chosen more often than those with higher sparsity, which increases the initial convergence rates. A mixed sampling strategy can be used to overcome the artefacts brought by importance sampling while keeping the fast convergence rate. We here present 2D simulation results which demonstrate the convergence properties of the proposed algorithms. 3D reconstruction results further verify the effectiveness of GCSGD in term of presenting visually acceptable reconstruction results.
\section{Open areas}

Simulation results have shown that when the group size is fixed, there is a range for parameter $b$ (or $\beta$) that guarantee the convergence of GCSGD. However, currently the  choice on $b$ is based on experience and trials, thus we want to develop a more systematic method, or an adaptive self-correction scheme for the determination of parameter $b$ or $\beta$. Besides, in more realistic tomographic imaging settings, with noisy observations and an inaccurate system matrix, simply using current method is often difficult to obtain high quality reconstructions. As a result, incorporating regularisation terms into GCSGD is another important direction for our future work. We are also interested in applying the proposed algorithm in reality dataset under a distributed network and explore the properties and differences of synchronous and asynchronous communication strategies. In our paper we have tested the convergence property when $\gamma<1$, which can be viewed as a form of asynchronous communication in a parallel computing architecture.  More systematic investigations of this will be included in future work.

\section*{Acknowledgment}

The authors would like to thank Professor Paul White who gave us positive feedbacks on the project. We also thank Dr Jordan Cheer who suggested us to improve our description on the simulation results part.

\ifCLASSOPTIONcaptionsoff
  \newpage
\fi


\begin{IEEEbiography}[{\includegraphics[width=1in,height=1.25in,clip,keepaspectratio]{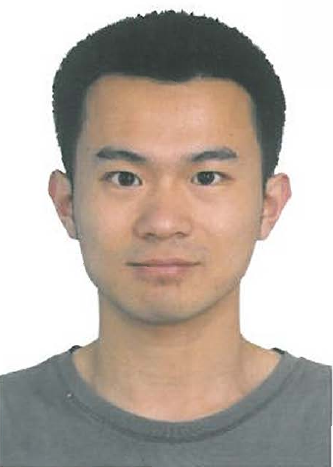}}]{Yushan Gao}
received the B.E. degree in optical engineering from Beijing Institute of Technology, Beijing, China, in 2010. He is currently pursing the Ph.D. degree with the Faculty of Engineering and Environment, University of Southampton, Southampton, UK. His research interests include image reconstruction and parallel computation.
\end{IEEEbiography}
\vskip 0pt plus -1fil
\begin{IEEEbiography}
[{\includegraphics[width=1in,height=1.25in,clip,keepaspectratio]{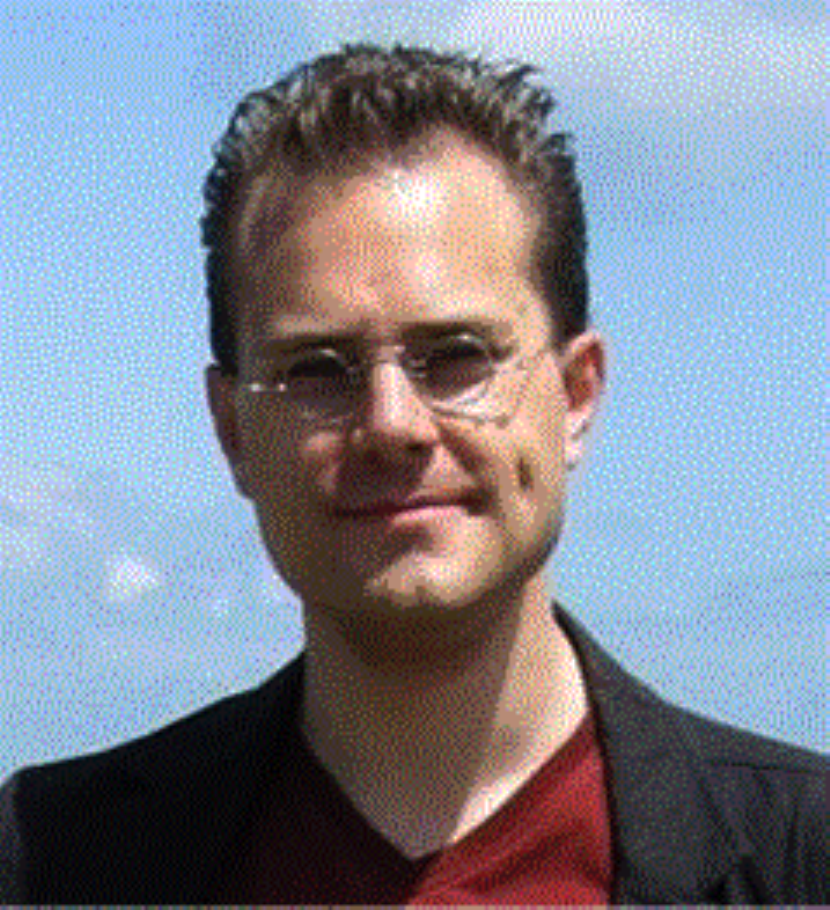}}]{Thomas Blumensath}(S'02-M'06)received the B.Sc. (Hons.) degree in music technology from Derby University, Derby, U.K., in 2002 and the Ph.D. degree in electronic engineering from Queen Mary, University of London, London, U.K., in 2006.  He specializes in statistical and mathematical methods for signal processing with a particular focus on sparse signal representations and their application to signal-processing problems.
\end{IEEEbiography}

\end{document}